\documentclass[12pt,A4widepaper]{article}
\usepackage{epsfig,amsmath,graphics}
\usepackage{graphics,graphicx}
\usepackage{tabularx}
\usepackage{amsfonts}
\usepackage{amsmath}

\def\ds{\displaystyle}

\newcommand{\beq}{\begin{equation}}
\newcommand{\eeq}{\end{equation}}
\newcommand{\lb}{\label}
\newcommand{\beqar}{\begin{eqnarray}}
\newcommand{\eeqar}{\end{eqnarray}}
\newcommand{\barr}{\begin{array}}
\newcommand{\earr}{\end{array}}

\newcommand{\derp}[2]{\ds{\frac {\partial #1}{\partial #2}}}

\def\scalp{\mbox{\boldmath$\, \cdot \, $}}

\def\XXint#1#2#3{{\setbox0=\hbox{$#1{#2#3}{\int}$}
     \vcenter{\hbox{$#2#3$}}\kern-.5\wd0}}

\def\bA{\mbox{\boldmath${\it A}$}}

\def\bB{\mbox{\boldmath${\it B}$}}
\def\bb{\mbox{\boldmath${\it b}$}}
\def\bC{\mbox{\boldmath${\it C}$}}

\def\bD{\mbox{\boldmath${\it D}$}}

\def\bE{\mbox{\boldmath${\it E}$}}
\def\be{\mbox{\boldmath${\it e}$}}
\def\bF{\mbox{\boldmath${\it F}$}}

\def\bL{\mbox{\boldmath${\it L}$}}

\def\bm{\mbox{\boldmath${\it m}$}}

\def\bn{\mbox{\boldmath${\it n}$}}
\def\b0{\mbox{\boldmath${\it 0}$}}

\def\bS{\mbox{\boldmath${\it S}$}}

\def\bt{\mbox{\boldmath${\it t}$}}

\def\bu{\mbox{\boldmath${\it u}$}}
\def\bv{\mbox{\boldmath${\it v}$}}

\def\bx{\mbox{\boldmath${\it x}$}}

\def\Id{\mbox{\boldmath${\it I}$}}

\def\bchi{\mbox{\boldmath${\chi}$}}

\def\btau{\mbox{\boldmath${\tau}$}}

\def\bSigma{\mbox{\boldmath${\Sigma}$}}

\def\tr{{\rm tr}}

\def\div{{\rm div}}
\def\Div{{\rm Div}}
\def\curl{{\rm curl}}
\def\Curl{{\rm Curl}}
\def\Grad{{\rm Grad}}
\def\grad{{\rm grad}}

\def\diag{{\rm diag}}

\def\matC{\mbox{\boldmath${\mathbb C}$}}

\def\matR{\mbox{\boldmath${\mathbb R}$}}

\def\capB{\mbox{\boldmath${\mathsf B}$}}

\def\E{\mbox{\boldmath${\mathcal E}$}}

\def\APL{{ Appl. Phys. Lett.\ }}

\def\IJES{{ Int. J. Eng. Sci.\ }}

\def\IJSS{{ Int. J. Solids Structures\ }}

\def\JAM{{ J. Appl. Mech.\ }}
\def\JAP{{ J. Appl. Phys.\ }}

\def\JMPS{{ J. Mech. Phys. Solids\ }}

\def\MOM{{ Mech. Materials\ }}

\def\PRSL{{ Proc. R. Soc. Lond.\ }}
\def\PRB{{ Phys. Rev. B\ }}

\def\salto#1#2{
[\mbox{\hspace{-#1em}}[#2]\mbox{\hspace{-#1em}}]}

\begin{document}

\title{The role of electrostriction on the stability of dielectric elastomer actuators}

\author{
   Massimiliano Gei$^*$, Stefania Colonnelli$^*$, Roberta Springhetti%
\footnote{Department of Civil, Environmental and Mechanical Engineering (DICAM),
University of Trento, Via Mesiano 77, I-38123 Trento, Italy; email:
massimiliano.gei@unitn.it; web-page: www.ing.unitn.it/$\sim$mgei.}
}

\maketitle

\begin{abstract}
In the field of soft dielectric elastomers, the notion \lq electrostriction' indicates the dependency of the permittivity on strain.
The present paper is aimed at investigating the effects of electrostriction onto the stability behaviour of homogeneous
electrically activated dielectric elastomer actuators. In particular, three objectives are pursued and achieved:
i) the description of the phenomenon within the general nonlinear theory of electroelasticity;
ii) the application of the recently proposed theory of bifurcation for electroelastic bodies in order to determine its role on the onset
of electromechanical and diffuse-mode instabilities in prestressed or prestretched dielectric layers;
iii) the analysis of band-localization instability in homogeneous dielectric elastomers.
Results for a typical soft acrylic elastomer show that electrostriction is responsible for an enhancement towards diffuse-mode instability,
while it represents a crucial property - necessarily to be taken into account - in order to provide a solution to the problem of
electromechanical band-localization, that can be interpreted as a possible reason of electric breakdown.
A comparison between the buckling stresses of a mechanical compressed slab and the electrically activated counterpart concludes the paper.
\end{abstract}

Keywords: Electroelasticity, Electroactive polymers, Smart materials, Electromechanical instability, buckling actuator

\section{Introduction}

Dielectric elastomer (DE) devices are electrically activated smart systems that possess mechanical properties similar to those of
natural muscles 
and therefore represent one of the most promising members within the class of the
artificial muscles (Bar-Cohen, 2001; Brochu and Pei, 2010).
Applications of these systems are common in the fields of mechatronics, aerospace, biomedical and energy engineering
as actuators, sensors, and energy harvesters (Carpi et al., 2008a). Their operating principle is based on the deformation of a
dielectric soft membrane induced by the electrostatic attraction forces arising between the charges placed on its opposite sides
(Pelrine et al., 1998, 2000);
such effect is proportional to the permittivity of the material, which unfortunately turns out to be very low for the typical materials in use
(e.g., silicones, acrylic elastomers) with relative dielectric constants $\epsilon_r$ amounting to {\bf a} few units.

While, on the one hand, research efforts are devoted to the design and realization of composite materials with significantly higher permittivities
to improve the electromechanical coupling (Zhang et al., 2002; Huang et al., 2004; deBotton et al., 2007; Carpi et al., 2008b, Molberg et al., 2010;
Bertoldi and Gei, 2011; Risse et al., 2012; Ponte Castaneda and Siboni, 2012; Tian et al., 2012; Gei et al., 2013),
on the other hand, the nonlinear theory of homogeneous soft dielectrics is still under way, in particular, special attention deserve the issues
associated with the different types of instability developing under operating conditions and those related to the intrinsic behaviour of the material, such as electrostriction and polarization saturation (Li et al., 2011a; Ask et al., 2012, 2013).

The aim of this paper is to give a contribution to the aspects just mentioned, by pursuing three main goals:

\begin{itemize}
\item to provide a framework accounting for electrostriction of soft DEs within the general nonlinear theory of electroelasticity.
As usual in the field of soft dielectrics, electrostriction is conceived as the dependency of the  relative permittivity of the material
on strain: this effect, experimentally observed (Wissler and Mazza, 2007; Li et al., 2011b), must be taken into account for modelling
purposes in view of the large deformations usually achieved. This phenomenon
has been theoretically addressed by Zhao and Suo (2008) who employed a simple model for its characterization;

\item to apply the general theory of bifurcation for electroelastic body proposed by Bertoldi and Gei (2011) to investigate ({\it i})
electromechanical instability in unconstrained specimens and ({\it ii}) diffuse-mode instabilities, including buckling-like and surface-like modes,
in prestretched dielectric layers.
In the aforementioned paper the focus was on layered composites (see also Nobili and Lanzoni, 2010, and Rudykh and deBotton, 2011),
while the current analyses are performed on homogeneous materials, for which the two types of bifurcation are obtained on the basis of a
common general criterion. Electromechanical instability on its own was extensively studied by methods developed by Zhao et al. (2007)
and De Tommasi et al. (2010), while De Tommasi et al. (2013) showed that an imperfection could trigger this instability
at a voltage much lower than that for a homogeneous specimen. Regarding the importance of diffuse modes, we mention that an Euler-like
instability is the activation mechanism of several types of buckling-like actuators (Carpi et al., 2008a; Vertechy et al., 2012);

\item to analyze band-localization instability in homogeneous DEs, in particular facing its relation with the constitutive properties of the solid.
The theory developed here extends to the electroelastic domain the well-known theory of localization of deformation in nonlinear elasticity,
where the existence of a localized solution  of the incremental problem -- concentrated within a narrow band -- is sought along the loading path
(Rice, 1973; Hill and Hutchinson, 1975; Bigoni and Dal Corso, 2008).
\end{itemize}

\begin{figure}[!t]
  \begin{center}
     \includegraphics[width= 14 cm]{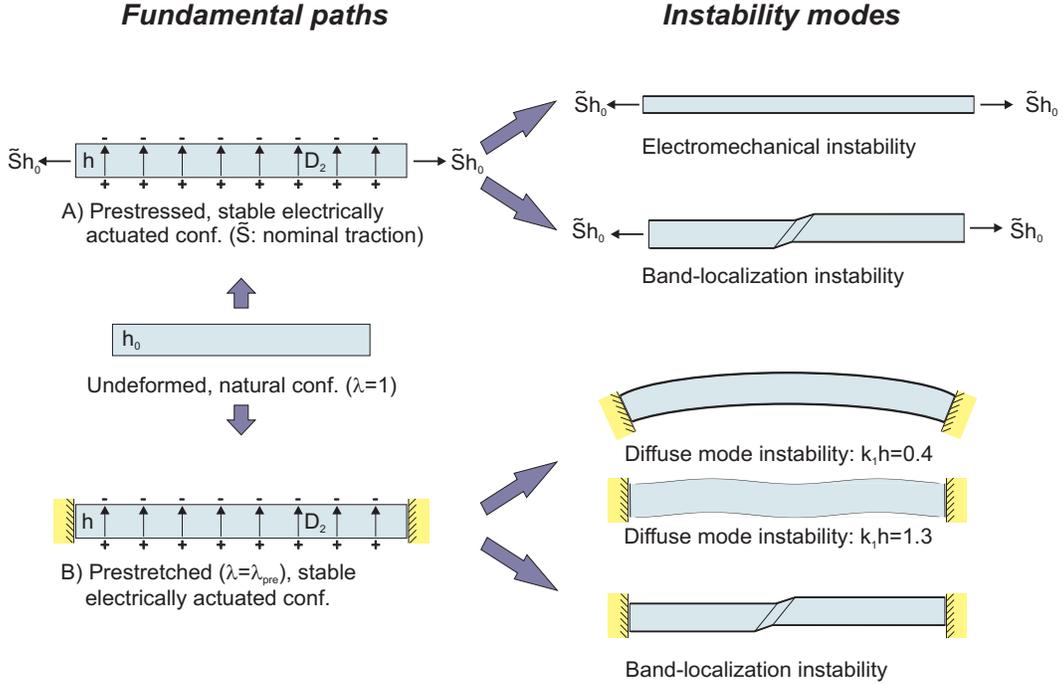}
      \caption{\footnotesize Sketch of the instabilities investigated in the paper for a soft dielectric layer subjected to two different
      electromechanical loading paths. A: the layer is electrically actuated with a constant longitudinal force --$\tilde{S}$ is the nominal
      traction--; B: the layer is first prestretched at a longitudinal stretch equal to $\lambda_{\rm pre}$ and then electrically actuated
      ($h_0$ and $h$ denote the initial and the reference thickness, respectively; $D_2$ represents the current electric displacement field).}
      \label{fig_introd}
   \end{center}
\end{figure}

General assumptions adopted throughout the paper involve plane-strain deformation and material incompressibility.
Fig. \ref{fig_introd} reports a sketch of the investigated instabilities relevant to the homogeneous loading
paths assumed for the layer that is always actuated by a given transverse electric displacement field: in the first (Path A), the prestressed specimen can freely expand under the electrical actuation,
while in the second (Path B), the layer is first mechanically prestretched and successively actuated.
The obtained results, well suited to a wide class of diffused acrylic elastomers, show that
electrostriction plays a fundamental role in the stability behaviour of the actuators.

The paper is organized into eight sections. Sects. 2 and 3 deal with the formulation of the finite and the linearized electroelastic models, respectively.
Sect. 4 introduces the considered electromechanical
loading paths, while in Sect. 5 the formulation of the general theory of electroelastic bifurcations introduced by Bertoldi and Gei (2011) is recalled and specialized to the plane-strain problem under study. The band-localization instability is discussed in Sect. 6, while all results and their interpretation are presented in Sect. 7. Finally, the conclusions are summarised in Sect. 8, while in Appendix A the components of the incremental moduli associated with the general free energy introduced in Sect. 2 are detailed.

\section{Large deformations and stress state for a soft dielectric body}

Here the theory of large-strain electroelasticity for a homogeneous isotropic hyperelastic body is briefly recalled, on the basis of the notion of total stress.
The reader is referred to McMeeking and Landis (2005), Dorfmann and Ogden (2005), Suo et al. (2008) and Bertoldi and Gei (2011) for further details.

A system in equilibrium under external electromechanical actions is considered, including an electroelastic body occupying
a region $B\in \matR^3$, whose points are denoted by $\bx$ and the surrounding space $B^{\rm sur}=\matR^3 \setminus B$.
Here the general case of the surrounding domain occupied by a different dielectric medium is briefly illustrated,
while in our reference problem we assume  $B^{\rm sur}$ corresponding to vacuum, in such case it will be denoted by $B^*$.
The stress-free configuration of the body $B^0$, whose points are
labelled $\bx^0$, can be identified, such that $\bx=\bchi(\bx^0)$,
where  $\bchi$ represents a given deformation and $\bF=\partial{\bchi}/\partial{\bx^0}$ denotes its gradient.
In the general case, the material configuration of the surrounding domain is
analogously denoted by ${B^0}^{\rm sur}$, while no reference configuration is introduced in the case of vacuum,
as the deformation gradient is not defined there.

\subsection{Field equations and boundary conditions}
\lb{sec_eq_gen}

Under the hypotheses previously introduced and assuming the absence of body forces and volume free charges,
the governing equations of the system  in the spatial description are:
     \beq
     \lb{DivtauDEeul}
     \div\,\btau = \b0,\ \ \ \ \btau^T=\btau,\ \ \ \ \div\, \bD=0,\ \ \ \ \curl\, \bE=\b0 \ \ \ \
     ({\rm in}\  B \cup B^{\rm sur}).
     \eeq
Here $\btau$ denotes the \lq total' stress, while $\bD$ and $\bE$ represent the electric displacement and the electric field respectively;
operators written with initial lower-case (upper-case) refer to variables defined in the present (reference) configuration.
Condition \eqref{DivtauDEeul}$_4$ states that $\bE$ is a conservative field, therefore it can be derived from a potential function $\phi(\bx)$, i.e.
$\bE=-\grad \phi(\bx)$, both in $B$ and $B^{\rm sur}$.

According to the considered set of boundary conditions, the charges are specified along the whole boundary $\partial B$, while
displacements and tractions prescriptions are enforced on disjoint parts of $\partial B$, denoted as $\partial B_v$
and $\partial B_t$, respectively, such that
$\partial B_v \cup \partial B_t=\partial B$ with $\partial B_v \cap \partial B_t= \emptyset$, namely
\beq
\lb{jump_eul}
\salto{0.1}{\bv}=\b0,\ \ \ \salto{0.1}{\btau} \bn=\bt\ \ \ ({\rm on}\ \partial B_t),
\ \ \ \bv=\tilde \bv \ \ \ ({\rm on}\ \partial B_v),
\eeq
$$
\salto{0.1}{\bD}\scalp \bn=-\omega,
\ \ \ \bn\times\salto{0.1}{\bE}=\b0\ \ \ ({\rm on}\ \partial B),
$$
where the jump operator defined on $\partial B$ corresponds to $\salto{0.1}{f}=f_{B}-f_{B^{\rm sur}}$,
${\bv}(\bx)$ denotes the finite displacement function with prescribed values $\tilde \bv$ on the restrained portion of the boundary $\partial B_v$,
$\bt$ and $\omega$ represent the assigned values of the tractions on the free boundary $\partial B_t$ and the surface charge density, respectively,
while $\bn$ is the current outward normal to $\partial B$.

The Lagrangian formulation of the above setting is also required, which is based on a back-mapping of the governing equations
(\ref{DivtauDEeul}) to the reference configuration $B^0\cup {B^0}^{\rm sur}$.
The variables involved are the first Piola-Kirchhoff total stress
\beq
\lb{stau}
\bS=J \btau \bF^{-T}
\eeq
and the material (or Lagrangian) version of the electric variables, i.e.
\beq
\lb{lagDE}
\bD^0=J \bF^{-1} \bD\ \ \  {\rm and}\ \ \  \bE^0=\bF^{T} \bE.
\eeq
In particular, under the same hypotheses, the field equations read now
     \beq
     \lb{DivtauDELag}
     \Div\,\bS = \b0,\ \ \ \ \bS\bF^T=\bF\bS^T,\ \ \ \ \Div\, \bD^0=0,\ \ \ \ \Curl\, \bE^0=\b0 \ \ \ \
     ({\rm in}\  B^0\cup {B^0}^{\rm sur}),
     \eeq
thus also the electric field $\bE^0$ proves to be conservative. The prescribed boundary conditions are analogous to those in (\ref{jump_eul})
for the corresponding Lagrangian variables
\beq
\lb{jump_lag}
\salto{0.1}{\bv^0}=\b0,\ \ \ \salto{0.1}{\bS} \bn^0=\bt^0\ \ \ ({\rm on}\ \partial B^0_t),
\ \ \ \bv^0=\tilde \bv^0 \ \ \ ({\rm on}\ \partial B^0_v),
\eeq
$$
\salto{0.1}{\bD^0}\scalp \bn^0=-\omega^0,
\ \ \ \bn^0\times\salto{0.1}{\bE^0}=\b0\ \ \ ({\rm on}\ \partial B^0),
$$
where $\bv^0(\bx^0)$ is the Lagrangian description of the finite displacement field, $\bt^0$, $\omega^0$ represent the nominal variables
of traction and surface charge density, respectively and $\bn^0$ is the unit vector normal to surface
$\partial B^0$.

Making reference back to the Eulerian formulation, when the surrounding space consists of vacuum (i.e. $B^{\rm sur} \equiv B^*$),
the stress in  $B^*$ reduces to Maxwell stress, here denoted by $\btau^*$,
$$
\btau^*=\epsilon_0 \left(\bE^*\otimes \bE^*-\frac{1}{2} (\bE^*\scalp \bE^*) \Id \right),
$$
where symbol * marks quantities evaluated in vacuum; moreover, electric displacement and electric field obey the law $\bD^*=\epsilon_0 \bE^*$, $\epsilon_0$
being the permittivity of vacuum ($\epsilon_0=8.85$ pF/m).
Field equations similar to \eqref{DivtauDEeul} can be stated, which are more explicative if the two domains, $B$ and $B^*$, are kept distinct:
     \beq
     \lb{DivtauDE}
     \div\,\btau = \b0,\ \ \ \ \btau^T=\btau,\ \ \ \ \div\, \bD=0,\ \ \ \ \curl\, \bE=\b0 \ \ \ \
     ({\rm in}\  B),
     \eeq
     \beq
     \lb{divcurlestar}
 \div\, \bE^*=0,\ \ \ \ \curl\, \bE^*=\b0 \ \ \ \ ({\rm in}\  B^*).
     \eeq
Here $\btau$ explicitly refers to the total stress in $B$, while on the basis of equations \eqref{divcurlestar} it can be easily shown
that Maxwell stress is divergence-free (Dorfmann and Ogden, 2010), therefore, being the symmetry of $\btau^*$ self-evident,
equations \eqref{DivtauDE} turn out to be formally valid also in vacuum and can be extended to the whole space $B \cup B^*$.
The associated boundary conditions are:
\beq
\lb{jumpboundary}
\btau \bn=\bt+\btau^* \bn\ \ \ \ ({\rm on}\  \partial B_t),\ \ \ \
\bv=\tilde\bv\ \ \ \ ({\rm on}\  \partial B_v),
\eeq
$$
\bD\scalp \bn=-\omega+\epsilon_0 \bE^* \scalp \bn,\ \ \ \
\bn \times (\bE-\bE^*)=\b0\ \ \ \ ({\rm on}\  \partial B).
$$

A Lagrangian version of the equations above can be provided for the dielectric body
     \beq
     \lb{DivtauDELag_vacuum}
     \Div\,\bS = \b0,\ \ \ \ \bS\bF^T=\bF\bS^T,\ \ \ \ \Div\, \bD^0=0,\ \ \ \ \Curl\, \bE^0=\b0 \ \ \ \
     ({\rm in}\  B^0),
     \eeq
unlike for vacuum, as no deformation and therefore no Lagrangian variables can be defined there, thus conditions \eqref{divcurlestar}
still should be enforced in vacuum. Analogously, the boundary conditions are expressed with reference to Lagrangian and Eulerian
variables inside the dielectric and vacuum, respectively
\begin{align}
\lb{jumpvaclag}
\bS \bn^0=\bt^0+J\btau^* \bF^{-T}_{\rm b} &\bn^0,\quad
\bD^0\scalp \bn^0=-\omega^0+\epsilon_0 J \bF^{-1}_{\rm b} \bE^* \scalp \bn^0,\\
&\bn^0\times \bE^0=\bn^0\times\bF^{T}_{\rm b} \bE^*,
\end{align}
where the notation $\bF_{\rm b}=\bF_{|\partial {B^0}}$ has been introduced.

\subsection{Constitutive equations}
\lb{sec_cost_eq}

We consider a conservative material, whose response can be described through a free-energy function $W=W(\bF,\bD^0)$ as

       \beq
        \lb{comp_const_eq_1}
        \bS=\derp{W}{\bF},\ \ \ \ \bE^0=\derp{W}{\bD^0},
        \eeq
or, in the case contemplated hereafter of an incompressible material (the dielectric elastomer is assumed to be incompressible,
being characterized by changes in shape typically much more significant than changes in volume), as
       \beq
        \lb{incomp_const_eq_1}
        \bS=\derp{W}{\bF}-p\, \bF^{-T},\ \ \ \ \bE^0=\derp{W}{\bD^0},
        \eeq
where $p$ represents an unknown hydrostatic pressure; the total stress $\btau$ and the current electric field $\bE$
can be easily obtained making use of eqs. \eqref{stau} and \eqref{lagDE}.

Isotropy requires that $W(\bF,\bD^0)$ be a function of the invariants of the right Cauchy-Green tensor $\bC=\bF^T \bF$,
(note that here $I_3=\det \bC=1$)
    \beq
    I_1=\tr\,\bC,\ \ \ \ I_2=\frac{1}{2}\left[(\tr\,\bC)^2-\tr\,\bC^2\right],
    \eeq
and of three additional invariants depending on $\bD^0$, namely
    \beq
    I_4=\bD^0\scalp\bD^0,\;\;\;I_5=\bD^0\scalp\bC\bD^0,\;\;\;I_6=\bD^0\scalp\bC^2\bD^0,
    \eeq
so that $W$ is a function of five independent scalars.
In particular, we will focus on the following form of the free energy
\beq
\lb{iso_psi}
W(I_i)=W_{\rm elas}(I_1,I_2)+\frac{1}{2\epsilon_0 \bar \epsilon_r}(\bar \gamma_0 I_4+\bar \gamma_1 I_5+\bar \gamma_2 I_6),
\eeq
where $\bar \epsilon_r$ is the relative dielectric constant of the material in the undeformed state ($\bF=\Id$) and $\bar\gamma_i\ (i=0,1,2)$
are dimensionless constants, such that $\sum_i \bar \gamma_i=1$.
In general, these coefficients can be, in turn, function of the same invariants, however  we do not take into account the more
general form here, as (\ref{iso_psi}) well captures the behaviour of ideal dielectrics
and electrostrictive materials, adopting constant coefficients.
Note that when electrostatic effects vanish (i.e. $I_4=I_5=I_6=0$), the free energy reduces to $W_{\rm elas}$.

The combination of eqs. (\ref{incomp_const_eq_1}) and \eqref{iso_psi}\footnote{
Note that function $W_{\rm elas}(I_1,I_2)$ has been assumed with no interconnection between invariants
$I_1$ and $I_2$, such that $\partial^2 W_{\rm elas}/\partial I_1\partial I_2=0$.
\vspace{2 mm}
}
after the derivatives of the invariants\footnote{
Derivatives of the invariants:
$$
\partial I_1/\partial \bF=2\bF,\quad \partial I_2/\partial \bF=2(I_1\bF-\bF\bC),$$
$$\partial I_4/\partial \bF=\b0,\quad \partial I_5/\partial \bF=2(\bF\bD^0)\otimes\bD^0,\quad \partial I_6/\partial \bF=2[(\bF\bD^0)\otimes(\bC\bD^0)+
(\bF\bC\bD^0)\otimes\bD^0],
$$
\vspace*{-.3cm}
$$
\partial I_4/\partial \bD^0=2\bD^0,\quad \partial I_5/\partial \bD^0=2 \bC\bD^0,\quad \partial I_6/\partial \bD^0=2 \bC^2\bD^0.
$$
}
in terms of $\bF$ and $\bD^0$ have been carried out and replaced, provides an explicit expression for the total stresses and electric fields.
In particular the Lagrangian variables are
\begin{align}
\nonumber
&\bS=-p \,\bF^{-T}+\mu\bigl[\bar \alpha_1 \bF-\bar \alpha_2 (I_1 \bF -\bF \bC)\bigr]+
\\
\lb{Siso_psi}
&\qquad +\frac{1}{\epsilon_0 \bar \epsilon_r}\bigl[\bar \gamma_1 \bF\bD^0 \otimes  \bD^0+\bar \gamma_2 (\bF\bD^0 \otimes  \bC\bD^0+\bF\bC\bD^0\otimes\bD^0)\bigr],
\\
\lb{E0iso_psi}
&\bE^0=(\E^0)^{-1} \bD^0,\qquad \mbox{where}\quad
(\E^0)^{-1}=\frac{1}{\epsilon_0 \bar \epsilon_r}(\bar\gamma_0\Id+\bar\gamma_1\bC+\bar\gamma_2 \bC^2),
\end{align}
being $\E^0$ the Lagrangian tensor of dielectric constants; the Eulerian variables are obtained through eqs. \eqref{stau} and \eqref{lagDE},
\begin{align}
\nonumber
&\btau=-p \, \Id+\mu\bigl[\bar \alpha_1  \bB-\bar\alpha_2 (I_1 \bB -\bB^2)\bigr]+
\\
\lb{tauiso_psi}
&\quad\,\,\,\,\, +\frac{1}{\epsilon_0 \bar \epsilon_r}\bigl[\bar \gamma_1 \bD \otimes  \bD+\bar \gamma_2 (\bD \otimes  \bB\bD+\bB\bD\otimes\bD)\bigr],
\\
\lb{eiso_psi}
&\bE=\E^{-1} \bD,
\qquad \mbox{where}\quad \E^{-1}=\frac{1}{\epsilon_0 \bar \epsilon_r}(\bar \gamma_0\bB^{-1}+\bar \gamma_1\Id+\bar \gamma_2 \bB),
\end{align}
with the definition of tensor $\E$, such as $\E^{-1}=\bF^{-T} (\E^0)^{-1} \bF^{-1}$, including the current dielectric constants.
In the equations above, where, if necessary, we will label with a superscript \lq el' (i.e. {\em electric}) the second row of eqs. (\ref{Siso_psi})
and (\ref{tauiso_psi}), $\mu \bar \alpha_1=2 \partial W/\partial I_1$ and $\mu \bar \alpha_2=-2 \partial W/\partial I_2$, being
$\mu$ the shear modulus in the undeformed state and $\bB$ represents the left Cauchy-Green strain tensor; note that the hydrostatic pressure $p$ is indeterminate and is evaluated enforcing the
boundary conditions of the electro-elastic boundary-value problem. Variable $\tilde p$ can be introduced as an alternative to $p$, such that
$
p=\tilde p+\bE \scalp \bD/2,
$
in accordance with what done by Zhao et al. (2007).

In general, as long as the hypotheses underlying expression \eqref{iso_psi} of the free energy are valid,
(\ref{tauiso_psi})-(\ref{eiso_psi}) provide the general response of an isotropic nonlinear electroelastic soft solid encompassing a
deformation-dependent electric -- {\em electrostrictive} -- response.
Relation (\ref{eiso_psi}) shows that the behaviour of an ideal dielectric, for which the permittivity is independent
of the current strain, i.e. $\bE=\bD/(\epsilon_0 \epsilon_r)$, is recovered imposing $\bar \gamma_0=\bar \gamma_2=0$ and $\bar \gamma_1=1$,
with $\bar \epsilon_r=\epsilon_r$. In this case, it is easy to notice that the association of the term multiplied by $\bar \gamma_1$ in (\ref{tauiso_psi})
and of the contribution $\bE \scalp \bD/2$ included in the hydrostatic pressure $p$ is recognizable as the internal Maxwell stress.
In general, while coefficient $\bar \gamma_0$ accounts for a purely dielectric contribution, $\bar \gamma_2$ couples electrostriction to the mechanical response, as evident in the total stress law.


As the soft dielectrics typically in use are mainly silicones and acrylic elastomers, two appropriate constitutive models are
Mooney-Rivlin and Gent (other models are likewise suitable, for instance Ogden and Arruda-Boyce models), respectively based on
the following forms of elastic energy:
\beq
W^{\rm MR}_{\rm elas}=\frac{\mu_1}{2}(I_1-3)-\frac{\mu_2}{2}(I_2-3),\ \ \ \mu=\mu_1-\mu_2 \ \ \ (\mu_2<0),
\eeq
\beq
\lb{Gentmu}
W^{\rm G}_{\rm elas}=-\frac{\mu}{2}J_m \ln\left[1-\frac{I_1-3}{J_m}\right]\!.
\eeq
Note that $J_m$ is the value taken by invariant $I_1-3$ when the molecular chains of the internal network of the polymer are fully stretched;
if the maximum stretch in a uniaxial test is taken to be 10, as suggested in Gent (1996), it turns out that $J_m=97.2$, providing $\lambda_{\rm max}=9.959$
in a plane-strain uniaxial test.
For the models above we have:
\begin{itemize}
\item Mooney-Rivlin model: $\quad\bar \alpha_1=\mu_1/\mu,\quad \bar \alpha_2=\mu_2/\mu$,
\item Gent model: $\qquad\qquad\,\,\quad\bar \alpha_1=\frac{J_m}{J_m-(I_1-3)},\ \ \ \bar \alpha_2=0$.
\end{itemize}

\subsection{Deformation-dependent permittivity: electrostriction}

\lb{electrosection}

Electrostriction is a term historically associated with the attitude of a material (polymeric or ceramic) to be deformed by
the application of an electric field.
In DEs, due to the large strains involved, this phenomenon concerns the variability of the dielectric permittivity
with the deformation (Zhao and Suo, 2008). Typical materials employed for DE actuators
are characterized by this property (Wissler and Mazza, 2007) and therefore it becomes important to investigate its effects towards the behaviour of such devices.
In particular, our goal is, firstly, to show that electrostriction is included in the constitutive model described above leading to equations
(\ref{Siso_psi}), (\ref{E0iso_psi}) or (\ref{tauiso_psi}), (\ref{eiso_psi}) and, secondly, to apply such equations in order to study the stability of DE actuators.

\begin{figure}[htbp]
  \begin{center}
      \includegraphics[width= 14 cm]{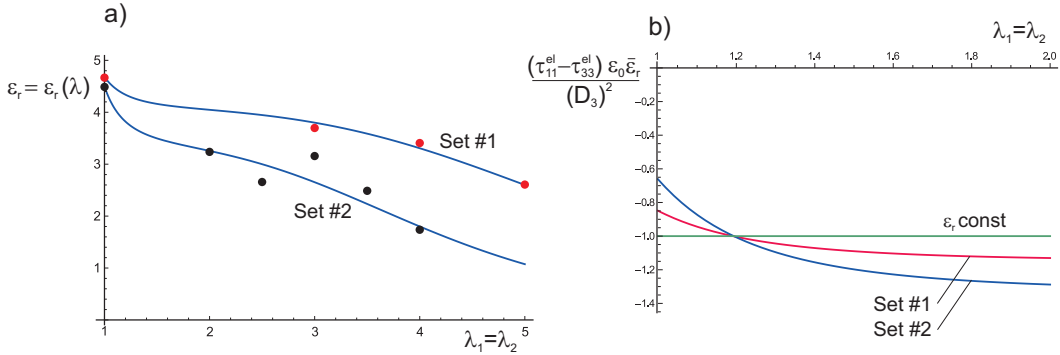}
      \caption{\footnotesize a) Fitting results based on expression (\ref{eiso_psi})$_2$ of the experimental data on electrostriction of
      3M VHB4910 provided by Wissler and Mazza (2007) and Li et al. (2011b) (see Table \ref{table_ele} for the values of
the parameters $\bar \gamma_i,\ i=0,1,2)$. b) Effect of the electrostriction on the biaxial \lq electric' stress-stretch response (\lq el' denotes the part of the stress depending on the electrostrictive parameters, see eq. (\ref{tauiso_psi})).}
      \label{fitting_elestriction}
   \end{center}
\end{figure}

The considered sets of parameters $\bar \gamma_i\ (i=0,1,2)$ have been assessed gathering data from the experimental tests performed by
Wissler and Mazza (2007) and Li et al. (2011b) on 3M VHB4910 equally biaxially prestretched films. For this purpose, formula (\ref{eiso_psi})$_2$
has been used to fit the experimental data, as depicted in Fig. \ref{fitting_elestriction}a where the in-plane stretches are equal ($\lambda_1=\lambda_2$),
providing the values reported in Table \ref{table_ele}.

The effect of electrostriction on the stress-strain behaviour of a soft dielectric layer
is illustrated in Fig. \ref{fitting_elestriction}b, where an equi-biaxial test ($\lambda_1=\lambda_2$) for an actuator activated imposing
an electric displacement field $D_3$ along the transverse direction is studied. There, the difference of the electric stress
$\tau_{11}^{\rm el}-\tau_{33}^{\rm el}$ ($\tau_{11}^{\rm el}=\tau_{22}^{\rm el}$) is sketched
in dimensionless form. For $\lambda_1=\lambda_2 >2$, the three curves remain almost parallel. It is clear that
the difference in the electromechanical response is appreciable even in the neighbourhood of the natural configuration ($\lambda_1=\lambda_2=1$).

\begin{table}
\centering
\begin{tabular}{lcccc} \hline
Set \# (Reference)& $\bar \epsilon_r$ & $\bar \gamma_0$ & $\bar \gamma_1$ & $\bar \gamma_2$ \\
\hline
1 {\footnotesize (Wissler and Mazza, 2007)}& 4.68 & 0.00104 & 1.14904 & $-0.15008$ \\
2 {\footnotesize (Li et al., 2011b)}& 4.5 & 0.00458 & 1.3298 & $-0.33438$ \\
\hline
\end{tabular}
\caption{Sets of electrostrictive parameters employed in the instability analyses.}
\lb{table_ele}
\end{table}

\section{Incremental electro-elastic boundary-value problem}
\lb{sec_incr}

The investigation of instabilities developing in dielectrics at large strains is carried out superposing incremental deformations
upon a given state of finite deformation (described in Section \ref{sec_eq_gen}).
Here we briefly introduce the topic, referring to Bertoldi and Gei (2011) for more details.

Again the general case is firstly presented, with the surrounding domain occupied by a different dielectric medium.
Let us assume a perturbation $\dot \bt\,\!^0$ and $\dot \omega^0$ of tractions and surface charges applied on $\partial B^0$ (henceforth a superposed dot will
denote the increment of the relevant quantity induced by the perturbation),
leading the system to a new equilibrium configuration. According to the Lagrangian formulation,
eqs. (\ref{DivtauDELag}) and (\ref{jump_lag}) hold true, as the body force density $\bb^0$ is unchanged.
The incremental problem is thus governed by the system
    \beq
    \lb{incrTotlagr}
    \Div\,\dot\bS=\b0,\;\;\;\;\Div\,\dot\bD\,\!^0=0,\;\;\;\;\Curl\,\dot\bE\,\!^0=\b0 \ \ \ \ ({\rm in}\  B^0\cup {B^0}^{\rm sur}),
    \eeq
with incremental jump conditions at the external boundary of the body taking the form
\beq
\lb{incr_jump_lag}
\salto{0.1}{\dot \bx}=\b0,\ \ \ \salto{0.1}{\dot \bS} \bn^0=\dot \bt\,\!^0\ \ \ ({\rm on}\ \partial B^0_t),
\ \ \ \dot \bx=\b0 \ \ \ ({\rm on}\ \partial B^0_v),
\eeq
$$
\salto{0.1}{\dot \bD\,\!^0}\scalp \bn^0=-\dot \omega^0,
\ \ \ \bn^0\times\salto{0.1}{\dot \bE\,\!^0}=\b0\ \ \ ({\rm on}\ \partial B^0),
$$
being $\dot \bx=\dot \bchi (\bx^0)$ the incremental deformation associated with the incremental deformation gradient $\dot \bF=\Grad \dot\bchi$.

Assuming that all incremental quantities are sufficiently small, the constitutive equations for a compressible medium (\ref{comp_const_eq_1}) can be linearized as
\beq
\lb{tot_sdot_compr}
\dot S_{iJ}=C^0_{iJkL} \dot{F}_{kL}+B^0_{iJL}\dot D\,\!^0_L,
\ \ \ \dot E\,\!^0_{M}=B^0_{iJM}\dot{F}_{iJ}+A^0_{ML}\dot D\,\!^0_L,
\eeq
where the components of the three electroelastic moduli tensors are given by
\beq
\lb{tot_cbaw}
C^0_{iJkL}=\frac{\partial^2 W}{\partial F_{iJ}\, \partial F_{kL}},\ \ \
B^0_{iJM}=\frac{\partial^2 W}{\partial F_{iJ}\, \partial D^0_M},\ \ \
A^0_{ML}=\frac{\partial^2 W}{\partial D^0_M\, \partial D^0_L}.
\eeq
From this definition, the following symmetries are derived:
\beq
\lb{symc0a0}
C^0_{iJkL}=C^0_{kLiJ},\ \ \ \ \ A^0_{ML}=A^0_{LM}.
\eeq

For incompressible materials ($\tr\dot\bF\bF^{-1}=0$, i.e. $\div\dot \bx=0$), the incremental total first Piola-Kirchhoff stress tensor is given by
\beq
\lb{tot_sdot}
\dot S_{iJ}=C^0_{iJkL} \dot{F}_{kL}+p\, F^{-1}_{Li}\dot{F}_{kL}F^{-1}_{Jk}-\dot p\, F^{-1}_{Ji}+B^0_{iJL}\dot D^0_L.
\eeq
The explicit expressions for the electroelastic moduli are detailed in Appendix A.

An updated Lagrangian formulation can be similarly provided for the incremental problem, based on the following field equations
    \beq
    \lb{incrToteul}
    \div\,\bSigma=\b0,\;\;\;\;\div\,\hat\bD=0,\;\;\;\;\curl\,\hat\bE=\b0 \ \ \ \ ({\rm in}\  B\cup {B}^{\rm sur}),
    \eeq
where $\bSigma=J^{-1}\dot\bS \bF^T$, $\hat\bD=J^{-1} \bF\dot\bD^0$ and $\hat\bE=\bF^{-T}\dot\bE^0$ correspond to incremental
updated variables obtained through a push-forward operation from the corresponding Lagrangian incremental variables (see \eqref{stau} and \eqref{lagDE}).
Identifying $\bu(\bx)=\dot \bx$, the associated incremental boundary conditions read
\beq
\lb{incr_jump_eul}
\salto{0.1}{\bu}=\b0,\ \ \ \salto{0.1}{\bSigma}\, \bn\, dA=\dot \bt\,\!^0 dA^0\ \ \ ({\rm on}\ \partial B_t),
\ \ \ \bu=\b0 \ \ \ ({\rm on}\ \partial B_v),
\eeq
$$
\salto{0.1}{\hat \bD}\scalp \bn dA=-\dot \omega^0 dA^0,
\ \ \ \bn\times\salto{0.1}{\hat \bE}=\b0\ \ \ ({\rm on}\ \partial B).
$$
Note that also the incremental electric field is conservative, both in Lagrangian and Eulerian formulation, what guarantees the existence of relevant incremental electrostatic potentials.

Also in the frame of the updated Lagrangian formulation, the incremental constitutive equations turn out to be linear and,
assuming $\bL=\grad \bu$, take the form
\beq
\lb{tot_sigmadot_compr}
\Sigma_{ir}=C_{irks} L_{ks}+B_{irk}\hat D_k,
\ \ \ \hat E_{i}=B_{kri}L_{kr}+A_{ik}\hat D_k.
\eeq
The expression of the incremental constitutive tensors is straightforwardly derivable from eqs. (\ref{tot_sdot_compr})
and (\ref{tot_sigmadot_compr}) through the definition of the updated Lagrangian variables, giving
\beq
\lb{tot_cbaw1}
C_{irks}=\frac{1}{J}\,C^0_{iJkL}{F}_{rJ}{F}_{sL},\ \ \
B_{irk}=B^0_{iJM}{F}_{rJ}{F}^{-1}_{Mk},\ \ \
A_{ik}=J\,A^0_{JM}{F}^{-1}_{Ji}{F}^{-1}_{Mk},
\eeq
where the following symmetry properties hold true:
\beq
\lb{symcba}
C_{irks}=C_{ksir},\ \ \ \ \ B_{irk}=B_{rik},\ \ \ \ \ A_{ik}=A_{ki}.
\eeq
Note that conditions (\ref{symcba})$_{1,3}$ are analogous to (\ref{symc0a0})$_{1,2}$, while (\ref{symcba})$_{2}$ can be established
by using the incremental form of the balance of angular momentum, also leading to condition
\beq
\lb{relC}
C_{iqkr}+\tau_{ir}\delta_{qk}=C_{qikr}+\tau_{qr}\delta_{ik}.
\eeq
In the case of an incompressible material, while symmetries (\ref{symcba}) still hold true, the updated version of the
incremental first Piola-Kirchhoff total stress tensor becomes
\beq
\lb{tot_sigma}
\Sigma_{ir}=C_{irks} L_{ks}+p{L}_{ri}-\dot p\, \delta_{ir}+B_{irk}\hat D_k
\eeq
while condition (\ref{relC}) turns into
\beq
\lb{relCinc}
C_{iqkr}+(\tau_{ir}+p\,\delta_{ir})\delta_{qk}=C_{qikr}+(\tau_{qr}+p\,\delta_{qr})\delta_{ik}.
\eeq
The detailed expressions of the moduli for the updated Lagrangian formulation is reported in Appendix A.

In the case the domain outside the solid is vacuum ($B^{\rm sur}\equiv B^*$), boundary conditions can be stated
as in Dorfmann and Ogden (2010) and Bertoldi and Gei (2011). Here we consider the case, relevant for practical applications, where
both surface tractions $\bt^0$ and surface charges $\omega^0$ are independent of the deformation (dead loading),
thus $\dot \bt^0=\bf {0}$ and $\dot \omega^0=0$, while the electric field in vacuum vanish, as in the space outside a parallel-plate capacitor
(by neglecting the edge effects). The consequence is that both the Maxwell stress $\btau^*$ and its increment $\dot \btau^*$, generally given as
$$
\dot \btau^*=\epsilon_0 \left[\dot \bE^*\otimes \bE^*+\bE^*\otimes \dot \bE^* - (\bE^{*}\scalp \dot \bE^{*}) \Id \right],
$$
vanish, while the increments of $\bD^*$ and $\bE^*$ (required in order to satisfy the incremental boundary conditions)
are simply related as $\dot \bD^*=\epsilon_0 \dot \bE^*$.
%
Therefore, also including the incompressibility of the dielectric,
the boundary conditions for the Lagrangian formulation of the incremental problem specialize as follows
\begin{align}
\lb{incr_jump_lag_vacuum}
{\dot \bS} \bn^0 =\b0,\quad\,\,
{\dot \bD\,\!^0}\scalp \bn^0=\epsilon_0( \bF^{-1}_{\rm b} \dot\bE^*)\scalp \bn^0,\quad\,\,
{\bn^0\times\dot \bE\,\!^0}=\bn^0\times \bF^{T}_{\rm b}\dot\bE^*,
\end{align}
while, with reference to updated Lagrangian variables, they read:
\begin{align}
\lb{incr_jump_eul_vacuum}
{\bSigma} \bn =\b0,\quad\,\,
{\hat \bD}\scalp \bn =\epsilon_0{\dot \bE^*}\scalp \bn,\quad\,\,
\bn\times{\hat \bE}=\bn\times{\dot \bE^*}.
\end{align}

Note that, owing to eq. (\ref{incrToteul})$_3$, the incremental electric variable $\dot\bE^*$ in $B^{\rm *}$ is profitably defined making
use of the incremental electrostatic potential in vacuum $\dot \phi^*(x_1, x_2)$ as
$\dot E^*_i=-\dot \phi^*_{,i}$; the fulfilment of condition (\ref{incrToteul})$_2$ in $B^*$ furthermore requires  that the potential function is harmonic:
\beq
\lb{harmonic}
\dot \phi^*_{,ii}=0.
\eeq

\section{Homogeneous fundamental paths: prestressed and prestretched layers}
\lb{fund_paths}

\begin{figure}[tbp]
  \begin{center}
      \includegraphics[width= 9 cm]{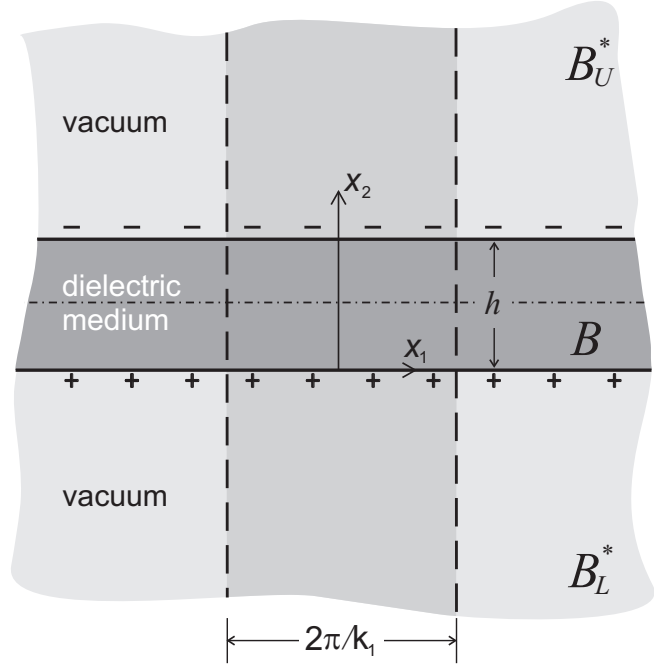}
      \caption{\footnotesize The general problem under study and the modular domain taken into account for the investigation of
              diffuse mode instability, according to the hypothesis of a periodic perturbation with wavelength equal to $2\pi/k_1$.}
      \label{FIGgeneral_problem}
   \end{center}
\end{figure}

Two states of electromechanical finite, plane-strain deformations are considered for a dielectric elastomer layer of initial thickness $h_0$,
as anticipated in Sect. 1 and depicted in Fig. \ref{fig_introd}. With reference to the current configuration, let $x_1$ and $x_2$
be the longitudinal and the transverse axes associated with the orthonormal basis $\{\be_1, \be_2 \}$ (being $\be_3$ the out-of-plane normal),
respectively, such that the boundaries of the layer correspond to $x_2=0,\,h$, as shown in Fig.  \ref{FIGgeneral_problem}.
We assume that the layer is infinitely wide and undergoes a homogeneous electric actuation aligned with direction $\be_2$ in the current configuration,
i.e. $\bD=D_2 \be_2$, with null external electric field, thus $\bE^{*}=\bD^{*}=\b0$. The deformation state is still homogeneous with deformation
gradient $\bF=\diag(\lambda,1/\lambda,1)$. The foreseen electrical activation can be achieved
applying a uniform distribution of opposite surface charges on the two boundaries, in this case the absolute value
of $D_2$ corresponds to the current charge density, see eq. \eqref{jump_eul}$_4$. The configuration can be also reached imposing a voltage
between two perfectly compliant electrodes placed on the two surfaces, but the bifurcation analysis requires an incremental problem where the voltage is the varied electrical quantity.

\subsection{Elongation under constant longitudinal force}

In this case (path A in Fig. \ref{fig_introd}) the actuator is stress free along direction $x_2$ and subjected to a constant force
$\tilde S h_0$ along the longitudinal direction, so that the nominal stress state is represented by
\beq
S_{11}=\tilde S,\ \ \ S_{22}=0,
\eeq
which provide the following implicit relation between $\lambda$ and $\bar D=D_2/\sqrt{\mu \epsilon_0 \bar \epsilon_r}$
\beq
\frac{\tilde S}{\mu} \lambda^2+ (\bar \alpha_1-\bar \alpha_2)\left(\frac{1}{\lambda}-\lambda^3\right)+\bar D^2
\left(\bar \gamma_1 \lambda+\frac{2 \bar \gamma_2}{\lambda}\right)=0.
\eeq
Graphical representations of this loading path are provided in Figs. \ref{loc_set1}a and \ref{loc_set2}a for electrostrictive Gent materials based on
the two sets of parameters mentioned above,
where the dimensionless electric displacement $\bar D$
is reported on the vertical axis of both plots.

\subsection{Pre-stretched specimen}

The so-called  path B depicted in Fig. \ref{fig_introd} is characterized by a total stress component along direction $x_2$ identically
vanishing throughout the solid, namely $\tau_{22}=0$, with the layer longitudinally prestretched at $\lambda=\lambda_{\rm pre}$
through the uniaxial tensile state of stress
\beq
\tau_{11}^{\rm pre}=\mu(\bar \alpha_1-\bar \alpha_2)\left(\lambda_{\rm pre}^2-\frac{1}{\lambda_{\rm pre}^2}\right).
\eeq
When an increasing electric displacement $D_2$ is subsequently superposed, the longitudinal stress changes as
\beq
\lb{tau_pre}
\frac{\tau_{11}}{\mu} =\frac{\tau_{11}^{\rm pre}}{\mu}-\bar D^2 \left(\bar \gamma_1+\frac{2 \bar \gamma_2}{\lambda_{\rm pre}^2}\right).
\eeq
The electric actuation yields a decrease in the longitudinal stress, therefore as shown in Figs. \ref{loc_set1}c and \ref{loc_set2}c
(representing eq. \eqref{tau_pre} for Gent materials with the two considered sets of parameters),
for increasing $D_2$, $\tau_{11}$ becomes negative involving a buckling-like (diffuse-mode) instability.
Condition $\tau_{11}=0$ is referred to as \lq null tension' threshold.

\section{Global instabilities of a soft dielectric elastomer}

Global equilibrium bifurcations for a generic electroelastic system consisting of two media, respectively occupying domains $B^0$ and
${B^0}^{\rm sur}$ with reference to a Lagrangian description
can be addressed referring to the general theory introduced by Bertoldi and Gei (2011).
Among this class of instabilities, for the electroelastic layer, we aim to investigate both electromechanical (pull-in)
and diffuse-mode bifurcations, involving the relevant cases of buckling-like and surface-like instabilities.

Along an electromechanical loading path, the existence of two distinct solutions of the incremental problem is admitted and the
fields generated as their difference, here denoted by symbol
$\Delta$ (e.g. $\Delta \dot \bchi=\dot \bchi^{(1)}-\dot \bchi^{(2)}$), are taken into account.
The difference fields can be regarded as the solution to a homogeneous incremental boundary-value problem
(no associated incremental body forces, tractions, volume free charges, surface charges),
thus an application of the principle of the virtual work in the material description requires that
\beq
\lb{pr_virt_mat}
\int_{B^0\cup {B^0}^{\rm sur}}\bigl[\Delta \dot\bS \cdot \Delta \dot\bF + \Delta \dot\bE^0 \cdot \Delta \dot\bD^0\bigr]\,dV^0=0
\eeq
for every set of admissible Lagrangian fields $\{\Delta \dot \bchi, \Delta \dot\bD^0, \Delta \dot\bS, \Delta \dot\bE^0\}$.
Note that the existence of the integrals on ${B^0}^{\rm sur}$ requires the decay at infinity of the fields involved.

Being both $\dot \bt^0$ and $\dot \omega^0$ null, the trivial pair $\{\dot \bchi^{(2)}, \dot\bD^{0(2)}\}=\bf{0}$ represents a
possible solution associated with the incremental boundary-value problem, consequently the difference fields reduce to the solution identified by
superscript $^{(1)}$ and equation (\ref{pr_virt_mat}) can be given the following form:
\beq
\lb{pr_virt_mat_1}
\int_{B^0\cup {B^0}^{\rm sur}}\Bigl[\dot\bS^{(1)} \cdot \dot\bF^{(1)} + \dot\bE^{0(1)} \cdot \dot\bD^{0(1)}\Bigr]\,dV^0=0.
\eeq
Therefore, denoting by $t\ (t\geq 0)$ the scalar loading parameter relevant to the principal equilibrium path,
a sufficient condition preventing the dielectric layer from the occurrence of a bifurcation is
\begin{align}
\lb{no_bifurc_mat}
\int_{B^0\cup {B^0}^{\rm sur}}\bigl[\dot\bS(t) \cdot \dot\bF + \dot\bE^0(t) \cdot \dot\bD^0\bigr]\,dV^0>0,
\end{align}
while a bifurcation takes place at $t=t_{cr}$ as soon as, for an admissible critical pair $\{\dot \bchi_{cr}, \dot\bD^0_{cr}\}$ --the
\emph{primary eigenmode}--, the functional becomes positive
semi-definite so that equation (\ref{pr_virt_mat_1}) becomes true, namely
\beq \lb{pr_virt_mat_cr} \int_{B^0\cup {B^0}^{\rm sur}}\bigl[\dot\bS_{cr}(t_{cr}) \cdot \dot\bF_{cr} +
\dot\bE_{cr}^0(t_{cr}) \cdot \dot\bD^0_{cr}\bigr]\,dV^0=0,
\eeq
with $\dot\bS_{cr}(t_{cr})$ and $\dot\bE_{cr} ^0(t_{cr})$ given by the
incremental constitutive equations.

The instability criterion defined in eq. (\ref{pr_virt_mat_cr}) according to the Lagrangian description can be easily given an updated
Lagrangian expression through a formal push-forward operation, namely
%
\begin{align}
\lb{no_bifurc_spat}
\int_{B\cup B^{\rm sur}}\bigl[\bSigma_{cr}(t_{cr}) \cdot \bL_{cr} + \hat\bE_{cr}(t_{cr}) \cdot \hat\bD_{cr}\bigr]\,dV=0,
\end{align}
for an admissible critical pair $\{\bu_{cr}, \hat\bD_{cr}\}$. Admissibility of $\bu_{cr}$ and $\hat\bD_{cr}$
requires the fulfilment of field and boundary conditions, namely the incompressibility constraint, $\div\,\bu_{cr}=0$, as well as eqs.
\eqref{incr_jump_eul}$_{1,3,4}$ and \eqref{incrToteul}$_2$.
Therefore, starting with an admissible critical pair $\{\bu_{cr}, \hat\bD_{cr}\}$ and using eqs.
\eqref{tot_sigmadot_compr} with the introduction of $\bL_{cr}=\grad \bu_{cr}$
to compute the corresponding incremental equilibrated total stress and curl-free electric field
$\bSigma_{cr}$ and $\hat\bE_{cr}$ (respectively satisfying field eqs. \eqref{incrToteul}$_1$ and \eqref{incrToteul}$_3$),
imposing the critical condition (\ref{no_bifurc_spat}) is equivalent
to the enforcement of the boundary conditions (\ref{incr_jump_eul})$_{2,5}$
in weak form, as can be easily shown making use of divergence and  Stokes' theorems.

When the surrounding medium is vacuum, condition (\ref{no_bifurc_spat}) takes the form
\begin{align}
\lb{no_bifurc_spatFIN}
\int_{B}\bigl[\bSigma_{cr} \cdot \bL_{cr} + \hat\bE_{cr} \cdot \hat\bD_{cr}\bigr]\,dV + \int_{B^{\rm *}}\dot\bE^*_{cr} \cdot \dot\bD^*_{cr}\,dV=0,
\end{align}
which can be simplified as
\begin{align}
\lb{energy_simplified}
\int_{B}\bigl[\bSigma_{cr} \cdot \bL_{cr} + \hat\bE_{cr} \cdot \hat\bD_{cr}\bigr]\,dV +
\epsilon_0 \int_{\partial B^*} \, \dot\phi^*_{cr}\, \grad\,\dot\phi^*_{cr}\cdot\bn \,dA=0,
\end{align}
through equation (\ref{harmonic}), entailing $\dot\bE^* \cdot \dot\bD^*  = \epsilon_0\, \div (\dot\phi^* \grad\dot\phi^*)$,
and subsequent application of the divergence theorem to the integral on $B^*$ in (\ref{no_bifurc_spatFIN}).
Condition (\ref{energy_simplified}) can be further simplified with the integral on $\partial B^*$ transported along $\partial B$, as
will be shown later for the problem under study.

\subsection{Electromechanical instability}

This bifurcation may arise when the body is deformed homogeneously as effect of dead-load tractions/charges applied
to its boundary, therefore homogeneous perturbation fields $\bL$, $\hat\bD$, and $\dot\phi^*$ are considered. Note that in this case
the surface integral in \eqref{energy_simplified} vanishes, being $\grad\,\dot\phi^*=0$, therefore, as a result of homogeneity,
the instability criterion requires that the argument of the volume integral in \eqref{energy_simplified} vanishes, namely,
for an incompressible material,
\beq
\lb{local_PD_diel_inc}
 \matC(t)  \bL \scalp \bL+p(t)\, \tr \bL^2+2 \capB(t) \hat\bD \scalp \bL+
\bA(t)  \hat\bD \scalp \hat\bD=0
\eeq
for at least a pair $\{\bL,\, \hat\bD\}=\{\bL_{cr},\, \hat\bD_{cr}\}\neq \b0$, with $\tr \bL=0$.

Note that, for the sake of conciseness, subscript '$cr$' has been omitted here and will be hereafter.
Therefore, bifurcation is predicted in correspondence to the loss of positive definiteness of the quadratic form in eq. (\ref{local_PD_diel_inc})
(see Gei et al., 2012, for the application of this criterion to a homogeneous actuators and the comparison with
the method based on the Hessian of the total energy).

\subsection{Diffuse-mode instability}
\lb{sect_diff_mode}

Diffuse modes, corresponding to a plane-strain inhomogeneous response of the layer
with wavelength given by $2\pi/k_1$ (being $k_1$ the wave-number of the perturbation), are investigated.
The extreme cases of {\em long-wavelength } ($k_1\rightarrow 0$) and {\em surface instability} ($k_1\rightarrow +\infty$), where
the critical modes are strongly localized in the vicinity of the surface, are considered.
Making reference to Fig. \ref{FIGgeneral_problem}, diffuse bifurcation modes are described
representing the set of admissible incremental fields in condition (\ref{energy_simplified}) as sinusoidal functions.

Considering the updated Lagrangian formulation, the incremental boundary-value problem can be written in scalar notation in the form:
\beq
\lb{incr_index_diel}
 \Sigma_{11,1}+\Sigma_{12,2}=0,\;\;\;\Sigma_{21,1}+\Sigma_{22,2}=0,\;\;\;\hat D_{1,1}+\hat D_{2,2}=0,\;\;\;\hat E_{1,2}-\hat E_{2,1}=0
\ \ \ \ ({\rm in}\  B),
\eeq
\beq
\lb{incr_index_vac}
\dot E^*_i=-\dot\phi^*_{,i}, \;\;\;\dot \phi^*_{,ii}=0\;\;\;
\ \ \ \ (i=1,2,\,\, {\rm in}\  B^*),
\eeq
\beq
\lb{incr_index_bc}
\Sigma_{12} =0,\quad\,\,\Sigma_{22} =0,\;\;\;
{\hat D}_2 =\epsilon_0{\dot E^*_2},\;\;\;
{\hat E}_1 ={\dot E^*_1}\ \ \ \ ({\rm along}\  \partial B).
\eeq
The periodic solution adopted inside layer $B$,
\begin{align}
\nonumber
&u_1(x_1,x_2)= Vs\, e^{s k_1 x_2} \cos k_1 x_1,\qquad
u_2(x_1,x_2)= V\, e^{s k_1 x_2}  \sin k_1 x_1,
\\
\lb{var_cos_sol}
&\hat{D}_1(x_1,x_2)= \delta s\, e^{s k_1 x_2} \cos k_1 x_1,\qquad
\hat{D}_2(x_1,x_2)= \delta\, e^{s k_1 x_2}  \sin k_1 x_1,\qquad \qquad
\\
\nonumber
&\,\dot{p}(x_1,x_2)= Q \, e^{s k_1 x_2}  \sin k_1 x_1,
\end{align}
guarantees that both fields $\bu$ and $\hat \bD$ are divergence-free, as required by  incompressibility
and eq. (\ref{incr_index_diel})$_3$ (note that in the case of a compressible dielectric,  condition $\div \bu=0$ would not subsist,
but simultaneously variable $\dot p$ would disappear).

In order to fulfil the remote decay conditions in the surrounding space, the relevant solution is expressed on the basis of the following
harmonic electric potentials inside each of the portions $B^{\rm *}_U$ and $B^{\rm *}_L$ in which $B^{\rm *}$ has been split
according to Fig. \ref{FIGgeneral_problem}:
\begin{itemize}
\item $\quad \dot\phi^*(x_1,x_2)=F_{U} \sin k_1 x_1 e^{-k_1 x_2}$ \,\, in $B^{\rm *}_U=\{\bx \in B^{\rm *},\, x_2 \ge h\}$,
\item $\quad \dot\phi^*(x_1,x_2)=F_{L} \sin k_1 x_1 e^{+k_1 x_2}$ \,\,\, in $B^{\rm *}_L=\{\bx \in B^{\rm *},\, x_2 \le 0\}$.
\end{itemize}
The interface jump condition (\ref{incr_index_bc})$_3$ at $x_2=0,h$ is easily satisfied through a convenient choice of constants $F_{U}$ and $F_{L}$.

When modes (\ref{var_cos_sol}) are plugged into constitutive eqs. (\ref{tot_sigmadot_compr})$_2$ and (\ref{tot_sigma}) and the resulting expressions
into conditions (\ref{incr_index_diel})$_{1,2,4}$, a homogeneous system of equations for amplitudes $V$, $\delta$, $Q$ is generated:
\begin{align}
\nonumber
\left[\begin{matrix} k_1 s (-C_{1111}+C_{1122}+s^2 C_{1212}+C_{1221})&s^2 B_{121}&-1\\
-k_1(C_{2121}+s^2 (C_{2112}+C_{2211}-C_{2222}))&s(-B_{211}+B_{222})&-s\\
k_1 s (s^2 B_{121}+ B_{211}-B_{222})&s^2 A_{11}-A_{22}&0\end{matrix}\right]
 \left[\begin{matrix}V\\\delta\\Q\end{matrix}\right]=\left[\begin{matrix}0\\0\\0\end{matrix}\right].
\end{align}
Thus a non trivial solution is only admissible when the associated matrix of coefficients is singular, i.e. when the
following cubic equation in $s^2$ is satisfied:
\begin{align}
\lb{eqs_diff_inst}
\Omega_6 s^6 + \Omega_4 s^4 + \Omega_2 s^2 + \Omega_0=0.
\end{align}
For the fundamental paths investigated throughout the paper, the expressions of the constitutive tensors $A_{ik}$, $B_{irk}$ and $C_{irks}$
reported in Appendix A and their symmetry properties (\ref{symcba}), the coefficients of (\ref{eqs_diff_inst}) can be given the following simplified expressions:
\begin{equation}
\begin{split}
\Omega_6&=-B^2_{121}+A_{11}C_{1212},\\
\Omega_4&=-2 B_{121}(B_{121}-B_{222})-A_{22}C_{1212}-A_{11}(C_{1111}-2C_{1122}-2C_{1221}+C_{2222}),\\
\Omega_2&=-(B_{121}-B_{222})^2+A_{11}C_{2121}+A_{22}(C_{1111}-2C_{1122}-2C_{1221}+C_{2222}),\\
\Omega_0&=-A_{22}C_{2121}.
\end{split}
\end{equation}
According to the nature of the six solutions $s_i$, different regimes can be identified and the general solution
inside $B$ is built by superposition:
\begin{align}
\nonumber
&u_1(x_1,x_2)=\sum_{i=1}^6 V_i s_i\,  e^{s_i k_1 x_2} \cos k_1 x_1,\qquad
u_2(x_1,x_2)= \sum_{i=1}^6 V_i\,  e^{s_i k_1 x_2} \sin k_1 x_1,
\\
\lb{var_cos_solSIX}
&\hat{D}_1(x_1,x_2)= \sum_{i=1}^6 \delta_i s_i \, e^{s_i k_1 x_2} \cos k_1 x_1,\qquad
\hat{D}_2(x_1,x_2)= \sum_{i=1}^6 \delta_i \, e^{s_i k_1 x_2} \sin k_1 x_1,
\\
\nonumber
&\dot{p}(x_1,x_2)= \sum_{i=1}^6 Q_i\,  e^{s_i k_1 x_2} \sin k_1 x_1.
\end{align}

The critical conditions are now determined introducing the latter expressions into the stability criterion,
eq. (\ref{energy_simplified}), which can be further simplified by taking into account the bounded modular domain highlighted in Fig. \ref{FIGgeneral_problem} as
\begin{align}
\lb{energy_simplified_modular}
\nonumber
\int_{-\frac{\pi}{k_1}}^{\frac{\pi}{k_1}}\int_{0}^{h}\bigl[\bSigma \cdot \bL + \hat\bE \cdot \hat\bD\bigr]\,dx_2 dx_1 +\\
-
\epsilon_0 \int_{-\frac{\pi}{k_1}}^{\frac{\pi}{k_1}} \, \Bigl[\dot\phi^*\,\grad\dot\phi^*&\cdot\bn\big|_{x_2=h}+
\dot\phi^*\,\grad\dot\phi^*\cdot\bn\big|_{x_2=0}\Bigr]\,dx_1=0;
\end{align}
here $\bn$ denotes the outward normal unit vector relevant to the specific boundary portion of $B$ (as in Sect. \ref{sec_eq_gen}).
Eq. (\ref{energy_simplified_modular}) stems from the remote decay conditions of the electric fields inside vacuum
and the periodic nature of the perturbation, allowing the integrals on $\partial B^*$ to vanish along the vertical
surfaces bounding the integration domain (corresponding to the dashed lines in Fig. \ref{FIGgeneral_problem}).
This procedure has been applied to the problem under study (the results will be presented in Sect. 7):
the primary eigenmodes so obtained have been shown to coincide with those evaluated on the basis of the procedure
illustrated in Bertoldi and Gei (2011), where all the boundary conditions (\ref{incr_jump_eul}) are enforced in strong form.

\section{A local instability of soft dielectric elastomers: band-localization}

A potential local instability mode arising in large-strain solid mechanics is band localization, where fields at bifurcation exhibit a
discontinuity across a narrow band of unknown inclination. The
condition for its onset along the homogeneous path (here reference will be made to the paths illustrated in Sect. 4) can be determined investigating
the admissible jumps of the incremental quantities across the interface between the band (superscript \lq b') and the rest of the solid (superscript \lq o').
\begin{figure}[tb]
  \begin{center}
     \includegraphics[width= 8 cm]{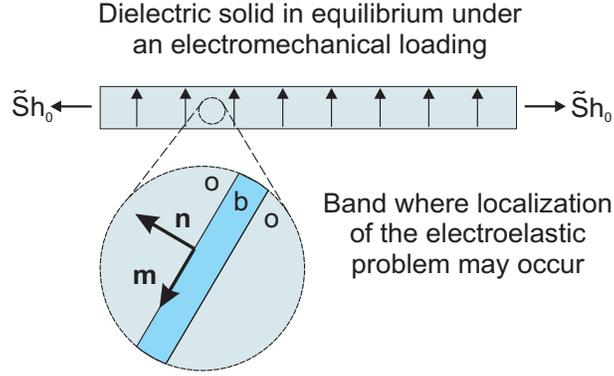}
      \caption{\footnotesize Band discontinuity in a homogeneously deformed dielectric elastomer.
      Band thickness is unpredictable on the basis of the proposed approach.}
      \label{fig_sketch_loc}
   \end{center}
\end{figure}
In the current configuration, let $\bn$ and $\bm$ denote two orthogonal unit vectors ($\bn\scalp \bm=0$),
normal to the band interface the first and aligned with it the latter, as depicted in Fig. \ref{fig_sketch_loc}%
\footnote{The vector $\bn$ used here must not be confused with the outward normal to $\partial B$ defined in Section 2.}.

Imagine that at the attainment of a threshold along the electro-mechanical loading path, $\bL^{\rm o}$ and $\hat \bD ^{\rm o}$
represent the uniform response of the solid to an incremental change in the boundary
conditions except inside the band, where the incremental displacement $\bu^{\rm b}$ is constant along
the planes $\bx \scalp \bn={\rm const}$ and the incremental electric displacement $\hat \bD ^{\rm b}$ is uniform.
Compatibility relationships across the interface, namely $(\bL^{\rm b}-\bL^{\rm o})\bm=\b0$
and continuity of the normal component of $\hat \bD$, respectively, require that
\beq
\lb{incr_jump_comp_upd}
\bL^{\rm b}=\bL^{\rm o}+\xi \bm \otimes \bn,\ \ \ \hat \bD^{\rm b}=\hat \bD^{\rm o}+\zeta \bm,
\eeq
where $\xi$ and $\zeta$ are real scalars representing mode amplitudes within the band; note the relative displacement
field in (\ref{incr_jump_comp_upd})$_1$, associated with the dyadic $\bm \otimes \bn$, that corresponds to an isochoric simple shear of amount $\xi$.
Both fields $\bL^{\rm b}$ and $\hat \bD^{\rm b}$ are required to satisfy field equations (\ref{incr_index_diel}) inside the band.

On the other hand, continuity of the increments of both traction and the tangential component of the electric field require
\beq
\lb{eq_salto_Sigma_E}
(\bSigma^{\rm b}-\bSigma^{\rm o})\bn=\b0,\ \ \ \ \hat  \bE^{\rm b}-\hat  \bE^{\rm o}=\beta \bn,
\eeq
where, again, $\beta$ is a real variable.
The use of (\ref{incr_jump_comp_upd}) in the constitutive equations and in (\ref{eq_salto_Sigma_E}) provides, in component form, respectively
\beq
\lb{salto_sigma_E_Q}
Q_{ik} m_k-\frac{1}{\xi}(\dot p^{\rm b}-\dot p^{\rm o}) n_i+ \bar \zeta B_{iqa} m_a n_q=0,
\eeq
$$
B_{iqa} m_i n_q+\bar \zeta A_{ab} m_b=\bar \beta n_a,
$$
where
$Q_{ik}=C_{iqkp} n_p n_q$, $\bar \zeta=\zeta/\xi$ and $\bar \beta=\beta/\xi$. Further manipulation of
(\ref{salto_sigma_E_Q}) yields
\beq
\lb{zetabar}
\bar \zeta=- \frac{B_{iqa} m_i n_q m_a}{A_{ab} m_a m_b},\ \ \ \ \bar \beta= B_{iqa} m_i n_q n_a+\bar \zeta A_{ab} n_a m_b,
\eeq
$$
\frac{1}{\xi}(\dot p^{\rm b}-\dot p^{\rm o})=Q_{ik} m_k n_i+\bar \zeta B_{iqa} n_i n_q m_a,
$$
as well as the condition for band localization, namely (assuming $A_{ab} m_a m_b\neq 0$)
\beq
\lb{loc_cond}
A_{ab} Q_{ik} m_a m_b m_i m_k-(B_{iqa} m_i n_q m_a)^2=0.
\eeq
Eq. (\ref{loc_cond}) clearly depends on the current state of finite-strain and on the normal to the band $\bn$ (the components
of $\bm$ can be easily substituted according to relation $m_r=e_{sr} n_s$, where $e_{12}=-e_{21}=1,\ e_{11}=e_{22}=0$).

For the fundamental paths under study, eq. (\ref{loc_cond}) explicitly becomes
\beq
\lb{eq_nu}
\Gamma_6 \nu^6+\Gamma_4 \nu^4+\Gamma_2 \nu^2+\Gamma_0=0,
\eeq
with the assumption $\nu=n_2/n_1$ ($n_1\neq0$) and the coefficients related to those in eq.  (\ref{eqs_diff_inst}) as:
\begin{equation}
\nonumber
\Gamma_6=-\Omega_6,\quad \Gamma_4=\Omega_4, \quad \Gamma_2=-\Omega_2, \quad \Gamma_0=\Omega_0.
\end{equation}

Band localization occurs when eq. (\ref{eq_nu}), which can be reduced to a cubic in the unknown $\nu^2$, admits a real solution $\nu^*$.
The real roots can be determined explicitly following Tartaglia-Cardano's theory
(valid for $\Gamma_6 \neq 0$;  when $\Gamma_6=0$, eq. (\ref{eq_nu}) becomes a biquadratic and the roots can be easily obtained).
According to the values taken by the discriminant
\begin{equation}
\label{discriminante}
\Delta= \frac{a^2}{4}+\frac{b^3}{27},
\end{equation}
where
\beq
a=-\frac13 \left(\frac{\Gamma_4}{\Gamma_6}\right)^2 +\frac{\Gamma_2}{\Gamma_6},\ \ \ \
b=\frac{2}{27} \left(\frac{\Gamma_4}{\Gamma_6}\right)^3 -\frac13 \frac{\Gamma_2 \Gamma_4}{\Gamma_6^2}+\frac{\Gamma_0}{\Gamma_6},
\eeq
two cases arise:
\begin{itemize}
\item  when $\Delta \geq 0$, eq. \eqref{eq_nu} has only one real root, i.e.
\begin{equation}
\lb{inst_delta_min0}
\nu^2=\sqrt[3]{-\frac{b}{2}+\sqrt{\Delta}}+\sqrt[3]{-\frac{b}{2}-\sqrt{\Delta}}- \frac{\Gamma_4}{3\Gamma_6};
\end{equation}
\item when $\Delta<0$, eq. \eqref{eq_nu} admits three real roots, namely
\beq
\nu^2_1=2 \sqrt{-\frac{a}{3}}\cos{\theta}- \frac{\Gamma_4}{3\Gamma_6},
\eeq
$$
\nu^2_2=2 \sqrt{-\frac{a}{3}} \cos{\left( \frac{\theta+2\pi}{3} \right)}- \frac{\Gamma_4}{3\Gamma_6}, \ \ \
\nu^2_3=2 \sqrt{-\frac{a}{3}} \cos{\left( \frac{\theta+4\pi}{3} \right)}- \frac{\Gamma_4}{3\Gamma_6},
$$
where $\theta=\arctan{(- 2\sqrt{-\Delta}/b)}$ if $b \leq0$ or $\theta=\pi+\arctan{( -2\sqrt{-\Delta}/b)}$ if $b>0$.
\end{itemize}

Along the principal path, the onset of band localization corresponds to the fulfilment of one of the following conditions:
i) $\Gamma_6 = 0$, ii) $\Gamma_0=0$, and iii) $\Delta=0$. The adoption of free-energy (\ref{iso_psi})
provides the following relation between $\lambda$ and $\bar D$ for case i)
\begin{equation}
\bar D=\sqrt{-\frac{  (\bar \alpha_1- \bar \alpha_2)(\bar \gamma_0 +\bar \gamma_1 \lambda^2 + \bar \gamma_2 \lambda^4)}
{\bar \gamma_2^2+\bar \gamma_0(\bar \gamma_1 \lambda^2+\bar \gamma_2 (2+\lambda^4))}},
\end{equation}
while for case ii) it gives
\begin{equation}
\bar D=\sqrt{-\frac{(\bar \alpha_1-\bar \alpha_2 )}{\bar \gamma_2} }.
\end{equation}
Case iii) is more involved, but a condition analogous to the previous ones can be easily determined from (\ref{discriminante}).

In any case, at the onset, the amplitude ratios $\bar \zeta$ and $\bar \beta$ defined in  eq. (\ref{zetabar}), become
\beq
\lb{zetabar_fund_paths}
\bar \zeta=-\frac{B_{121}\nu^3 +(B_{222}-B_{121})\nu}{A_{11}\nu^2+A_{22}}n_1 ,
\eeq
\beq
\lb{alphabar_fund_paths}
\bar \beta= B_{121}+(B_{222}-B_{121})\nu^2+\bar \zeta (A_{22}-A_{11})\nu.
\eeq

\section{Results}

\subsection{Diffuse-mode instability}

\begin{figure}[!t]
\begin{center}
\includegraphics{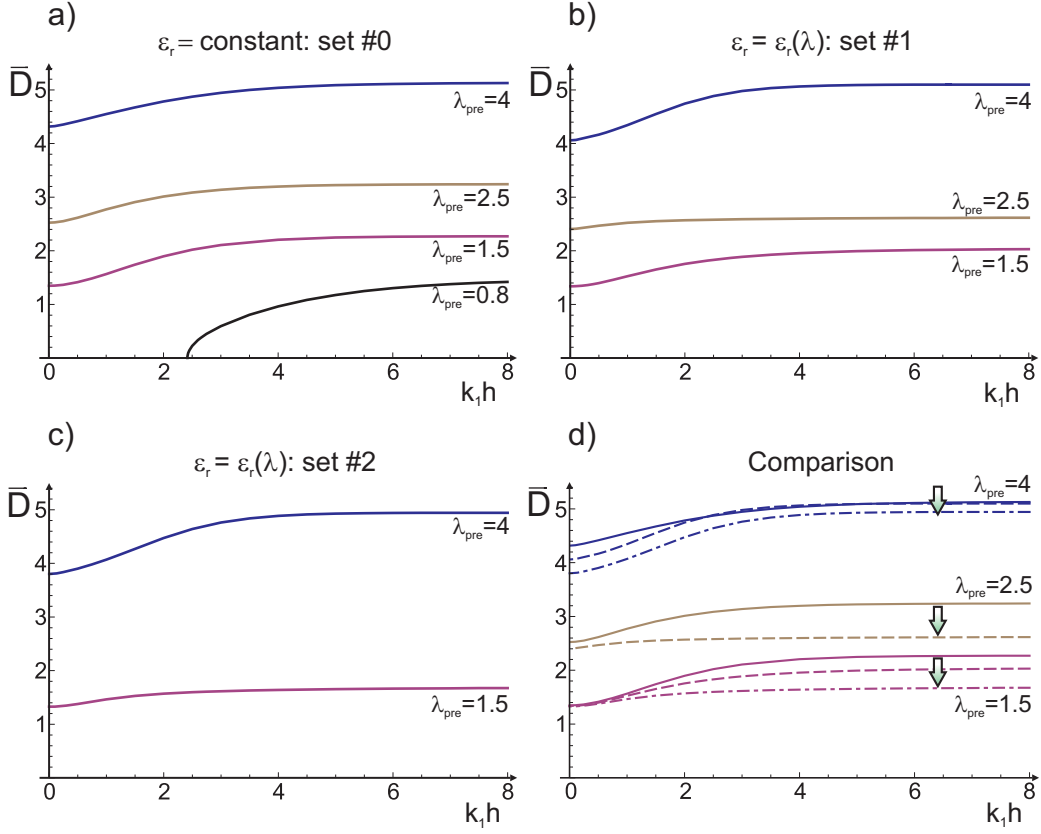}
\caption{\footnotesize {Diffuse instability modes for a prestretched actuator in plane strain (for a Gent material model).
Parts a), b), c): critical dimensionless electric displacement $\bar{D}$ vs dimensionless wavenumber $k_1 h$ for constant $\epsilon_r$ and for
the two sets of electrostrictive parameters considered in Table \ref{table_ele} at different values of the applied
prestretch $\lambda_{\textrm{pre}}$. The comparison reported in d) shows that electrostriction significantly lowers the critical
$\bar{D}$. In c) the case $\lambda_{\textrm{pre}}=2.5$ is not reported as it lies within a band-localization range (see Fig. \ref{loc_set2}).}}
\label{fig_diffuse}
\end{center}
\end{figure}

Diffuse-mode instability results are depicted in Fig. \ref{fig_diffuse} for
a prestretched specimen with different $\lambda_{\textrm{pre}}$ (path B in Fig. \ref{fig_introd}),
on the basis of an extended Gent electroelastic free energy (\ref{Gentmu}), characterized by
different sets of electrostrictive parameters (see Table \ref{table_ele}).
Both symmetric and antisymmetric modes (with respect to the symmetry axis of the layer, see Fig. \ref{FIGgeneral_problem}; see
also Bigoni and Gei, 2001) have been carefully checked and the critical conditions
have always been proved to correspond to antisymmetric modes.

In all the plots the dimensionless electric displacement $\bar D$, acting as
\lq electrical' loading parameter, is plotted as a function of the dimensionless wavenumber
$k_1 h$; note that the limit $k_1 h \rightarrow \infty$ denotes a surface-like mode\footnote{At high frequencies, the critical
$\bar D$ for symmetric and antisymmetric modes converges to the same value.},
while low values of $k_1 h$ correspond to buckling-like modes.
The latter case is well depicted by the graphical sketch of modes $k_1 h=0.4, 1.3$ in Fig. \ref{fig_introd}.

The effects of electrostriction onto the critical electric displacement at bifurcation are represented
in part d) of Fig. \ref{fig_diffuse}, where the comparison between the computations displayed in parts a), b), c) is reported.
In general, a high degree of electrostriction entails more evident reductions in the critical electric actuation
(specially for $\lambda_{\rm pre}=1.5, 2.5$ that are levels of prestretch important in the applications).
This can be justified observing that instability occurs when the axial stress $\tau_{11}$, tensile just after the prestretching,
becomes compressive. As can be observed comparing two paths at the same $\lambda_{\rm pre}$ in parts c) of Figs. \ref{loc_set1} and \ref{loc_set2},
at high electrostriction this event takes place for a slightly lower $\bar D$.

It is worth highlighting that experimental results on electrostriction are only available for stretched membranes (see Sect. \ref{electrosection}),
therefore the estimated values of parameters $\bar \gamma_i$ well interpolate the behaviour for $\lambda_{\textrm{pre}}>1$, while for $\lambda_{\textrm{pre}}<1$
we have noticed that the consequent dielectric constant $\epsilon_r$ is far from reasonable values. As a consequence, for $\lambda_{\textrm{pre}}=0.8$
only the curve for constant $\epsilon_r$ has been sketched in Fig. \ref{fig_diffuse} a).
For Set $\# 2$, calculations show that at a prestretch  $\lambda_{\rm pre}=2.5$ the specimen is
in the conditions where, along the electromechanical deformation, band-localization instability first takes place
and the previous homogenous response of the layer is lost (see below): for this reason the curve for $\lambda_{\rm pre}=2.5$ has not been illustrated.

\subsection{Band-localization instability}

Band-localization instability analysis for homogeneously deformed actuators is reported in Figs. \ref{loc_set1}, \ref{loc_set2} for an extended
Gent free-energy function with set of parameters $\#1$ and $\#2$, respectively, for both fundamental paths introduced in Sect. \ref{fund_paths}.
In a) the actuator is prestressed with a given nominal traction $\tilde S$,
following a nonlinear electroelastic deformation corresponding to path A, while in b) and c) the specimen is prestretched
at $\lambda=\lambda_{\textrm{pre}}$ and then actuated (path B), as for the analysis of diffuse modes.
In a), and c) dashed portions of the loading path curves (bounded by circles) denote ranges where band localization occurs.
Even though the current analysis allows to predict only the onset of such instability, while nothing can be said about the evolution of the band,
we note that in electroelasticity stable homogeneous nonlinear deformations are also possible beyond the theoretical emergence of the band,
suggesting that the range of instability can be crossed in some way, in order to reach the stable
path anew (the same applies to electroelastic deformations where the actuator deforms biaxially --computations not reported).
Comparison with experiments is difficult, as we are not aware of papers dealing with electroelastic band-localization instability
and this article provides the first theoretical analysis on the topic.
The following comments must be added to clarify the key points of our investigation:

\begin{figure}[!htbp]
  \begin{center}
     \includegraphics[width= 7 cm]{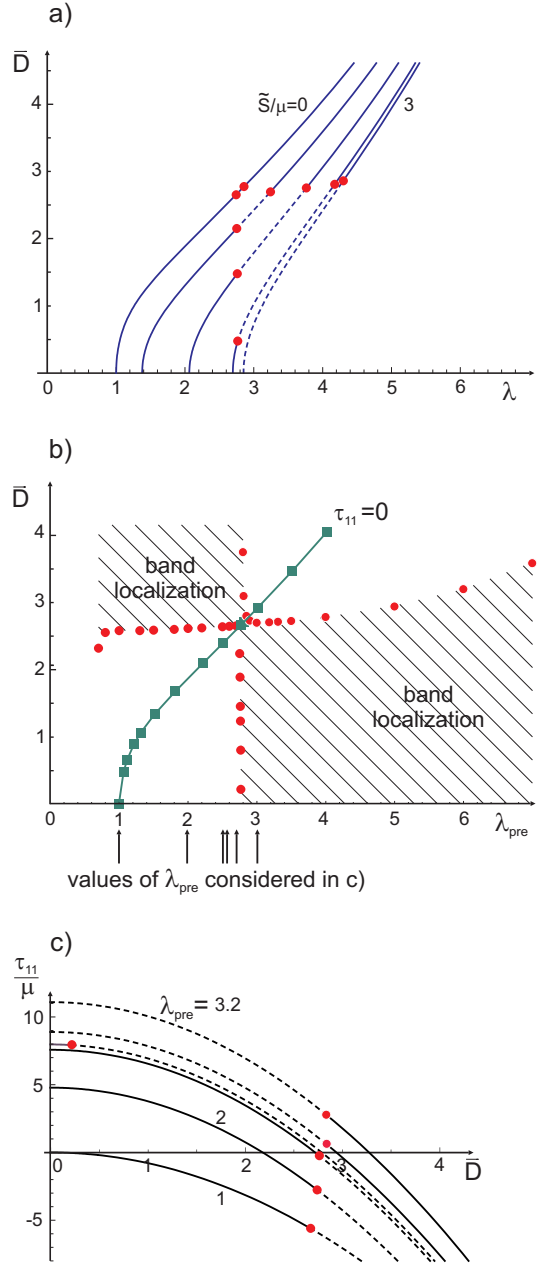}
\caption{\footnotesize Band-localization instability analysis for an electrically actuated DE layer in plane strain (Gent material model,
set of parameter $\# 1$, see Table 1).
a): plot for prestressed actuators with different $\tilde S/\mu$ (path A, in ascending order $\tilde S/\mu=0, 1, 2 , 2.8, 3$); dashed
lines indicate ranges where instability occurs (bounded by circles).
b), c): results for actuators initially prestretched at $\lambda=\lambda_{\rm pre}$ (path B). In particular, b): instability region in the
$\lambda_{\rm pre}$--$\bar D$ diagram -- the line marked by squares corresponds to $\tau_{11}=0$ (\lq null tension' threshold:
beyond this line the specimen is compressed); c): dimensionless longitudinal stress ($\tau_{11}/\mu$) and localization ranges 
in terms of electrical actuation $\bar D$.
}
    \lb{loc_set1}
  \end{center}
\end{figure}

\begin{figure}[!htbp]
  \begin{center}
      \includegraphics[width= 7 cm]{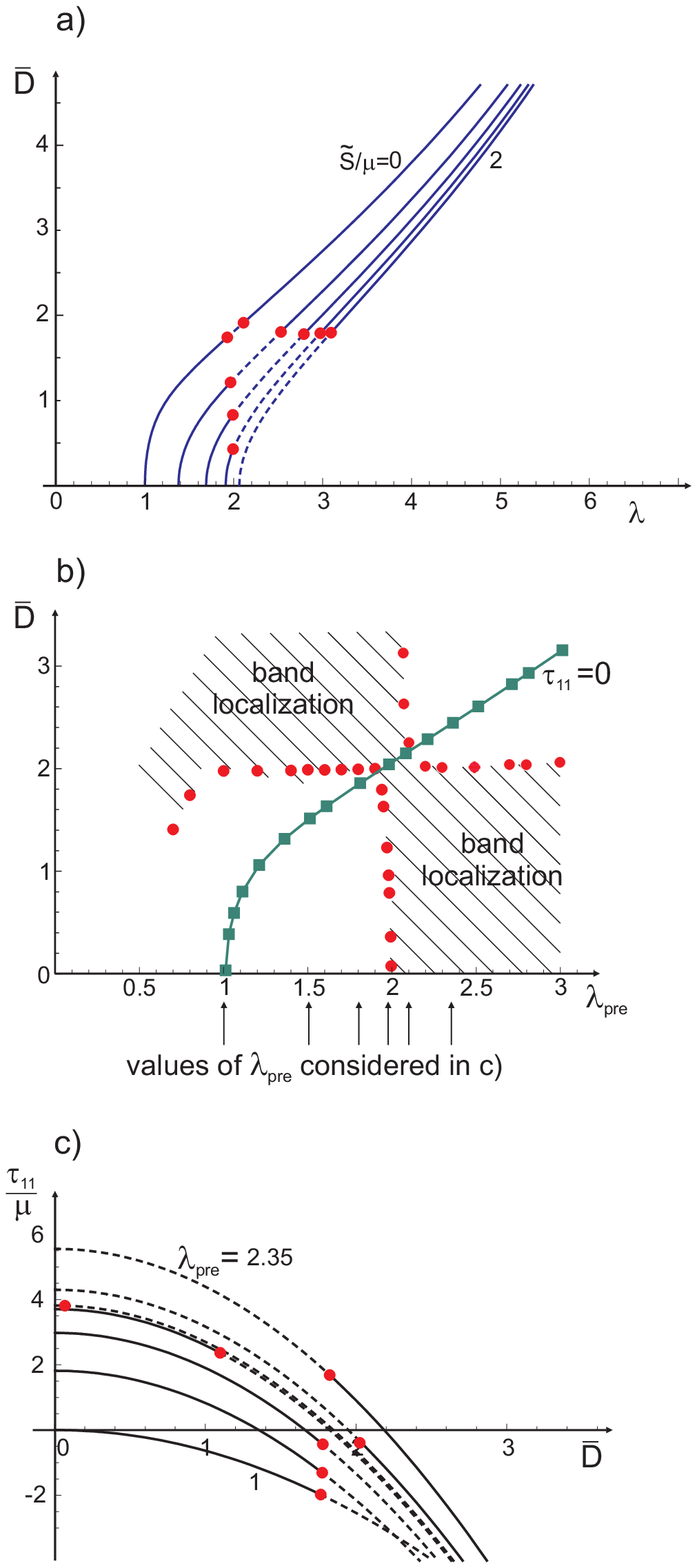}
\caption{\footnotesize Band-localization instability analysis for an electrically actuated DE layer in plane strain (Gent material,
set of parameter $\# 2$, see Table 1). a): plot for prestressed actuators with different $\tilde S/\mu$
(path A, in ascending order $\tilde S/\mu=0, 1, 1.5 , 1.8, 2$); dashed lines indicate ranges where instability occurs (bounded by circles).
b), c): results for actuators initially prestretched at $\lambda=\lambda_{\rm pre}$ (path B). In particular, b): instability region in the
$\lambda_{\rm pre}$--$\bar D$ diagram -- the line marked by squares corresponds to $\tau_{11}=0$ (\lq null tension' threshold:
beyond this line the specimen is compressed); c): dimensionless longitudinal stress ($\tau_{11}/\mu$) and localization ranges 
in terms of electrical actuation $\bar D$.
}
    \lb{loc_set2}
  \end{center}
\end{figure}

{\it i}) the onset of localization is strongly dependent on electrostriction. For a material with \emph{deformation independent permittivity} (i.e.
$\epsilon_r=\bar \epsilon_r$), no localization is predicted on the basis of eq. (\ref{eq_nu}).
Therefore, to detect the emergence of a band, accurate experiments must be carried out in order to carefully measure and identify the
electrostrictive properties of the specimen. It is worth pointing out  that Gent elastic model does not exhibit localization under
pure mechanical loadings, thus the instabilities observed here are genuine electromechanical effects;

{\it ii}) polynomial (\ref{eq_nu}) is obtained assuming $\hat \bD$ as the independent electric incremental variable, what physically corresponds to
perturb the surface charge applied on the layer boundaries\footnote{The technique assuming the control of the charge on the layer boundaries is less
common than the one based on the control of the voltage applied by the electrodes, nevertheless it is possible and has been successfully
employed by Keplinger et al. (2010).}.
Alternatively, a similar analysis can be carried out perturbing the voltage at the electrodes, therefore
choosing $\hat \bE$ as the primary variable. Even though the governing equations are the same,
the analogous of (\ref{eq_nu}) may exhibit properties being substantially
different from those of (\ref{eq_nu}), as the relevant constitutive equations accounting for the coupling differ from those
presented in Sect. \ref{sec_incr}. This analysis is out of the scope of the present paper and will be developed elsewhere;

{\it iii}) a failure mode experimentally observed in DE actuators is electric breakdown: when the electric field
inside the solid reaches a material-dependent threshold, the dielectric becomes conductive, with a discharge crossing the solid and inducing a strong localized
damage to the actuator.
We suggest that electric breakdown can be induced by a band-localization instability. Indeed, at the onset of this instability and for both fundamental paths,
the band inclination predicted on the basis of our analysis has always proven to be orthogonal to the direction of the electric field
(i.e. $\nu\rightarrow \infty$, while coordinate $x_2$, where the band develops, remains unknown). Through relationships (\ref{zetabar})
we can estimate the incremental fields inside the band
by setting the amplitude $\xi$; this has been done for the case $\tilde{S}/\mu=0$ in Fig. \ref{loc_set1}a, showing that the increment of the electric field
$\hat{\bE}^{\rm b}$ inside the band is almost six times larger then that outside (i.e., $\hat{\bE}^{\rm o})$, with a strong localized behaviour
of the incremental electric field. This can obviously match with micromechanical issues in order to promote electric breakdown.
From the previous considerations, it appears evident that the development of a band in a real sample represents something uncertain,
requiring additional investigations, both experimental and theoretical.
As for the latter aspect, it could be relevant to adopt a microelectromechanical model, in order to follow the evolution of the band and check the stability of the
predicted shear bands.

Coming back to Fig. \ref{loc_set1} (relevant to the set of parameters $\#1$),
for both fundamental paths it appears clear that $\lambda \approx 2.76$ provides a theoretical critical threshold.
As anticipated, this limit strongly depends on the degree of electrostriction, as shown in Fig. \ref{loc_set2}
for set $\#2$, where the same limit drops to approximately 1.97.
For prestretched actuators similar considerations apply, as depicted in parts b) and c) of Figs. \ref{loc_set1} and \ref{loc_set2}.
In parts b), in addition to the regions where localization represents the theoretically critical condition, the line corresponding to a null
longitudinal stress ($\tau_{11}=0$, \lq null tension' threshold, as an effect of electric actuation $\bar D$), is also reported, as typical
devices must operate under a tensile stress state in order to avoid buckling instability. Therefore, only the points
at the right-hand side of the line $\tau_{11}=0$ correspond to sensible configurations for real actuators.
The arrows below the horizontal axis in parts b) of both figures
(ranging from $\lambda_{\rm pre}=1$ to $\lambda_{\rm pre}=3.2$ in Fig. \ref{loc_set1} and between 1 and 2.35 in Fig. \ref{loc_set2})
 refer to the loading paths indicated in part c).

\subsection{Buckling instability: mechanically compressed vs prestretched and electrically activated slabs}

\begin{figure}[!t]
\begin{center}
\includegraphics[width= 16 cm]{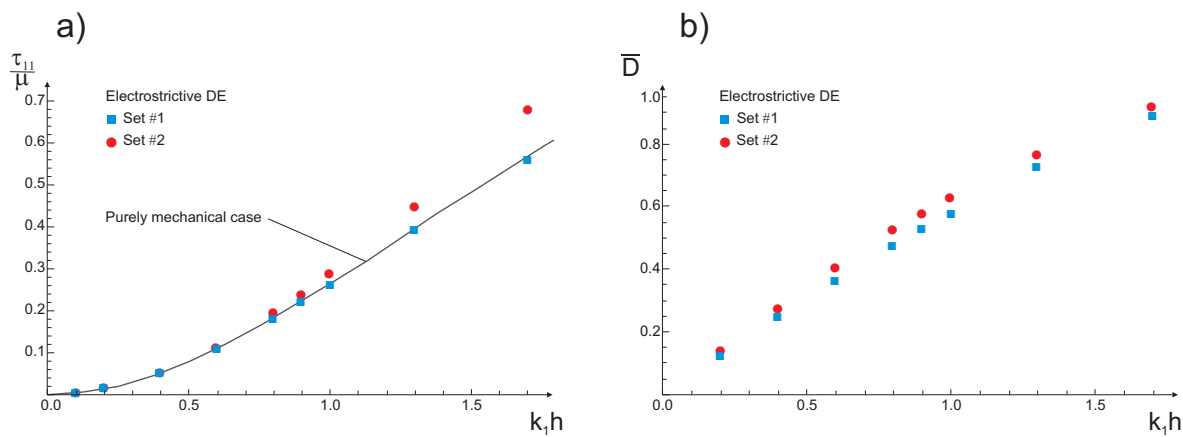}
\caption{\footnotesize {Buckling instability analysis in plane strain for a purely mechanically compressed (Mooney-Rivlin material)
and an electrically actuated DE layer ($\lambda_{\rm pre}=1$, Mooney-Rivlin material, sets of parameters $\# 1, 2$, see Table 1).
a): dimensionless longitudinal true stress for the mechanically compressed (continuous line: as $\bar D=0$, $\tau_{11}$ reduces to the Cauchy stress)
and the electromechanically activated DE (scattered points corresponding to the sets of parameters indicated in the legend)
vs. the dimensionless wavenumber $k_1 h$. b): dimensionless electric displacement $\bar D$ at instability vs. $k_1 h$
for the electromechanically activated DE.}}
\label{buckling}
\end{center}
\end{figure}

Even though the bodies investigated in this paper are electrically activated, the diffuse-mode instabilities analysed in Sect.
\ref{sect_diff_mode} are essentially driven  by the induced
compressive longitudinal stress arising as a reaction to the imposed boundary constraints. Therefore, it seems interesting to address
the following question: which longitudinal stresses are responsible for a common buckling mode in two identical silicone-like specimens,
mechanically loaded the former and electrically activated the latter?
In order to provide an answer, an isotropic thin layer with constitutive behaviour
described by a Mooney-Rivlin elastic energy is taken into account, for which two different plane strain fundamental paths are considered
(same geometry as in Fig. \ref{FIGgeneral_problem}):
{\em i}) a purely mechanical longitudinal compression, i.e. $\lambda_{\rm pre}<1$ (but remaining in the neighbourhood of 1) with $\bar D=0$;
{\em ii}) an electric actuation ($\bar D>0$) as in the path B described previously, with $\lambda_{\rm pre}=1$. The bifurcation analysis of the first problem is well-known (see Biot, 1965)
and is summarized here as the continuous curve illustrated in Fig. \ref{buckling}, representing the compressive longitudinal true stress at the onset of instability
(actually, when no electric effects are present, the total stress $\tau_{11}$ reduces to the Cauchy stress).
The second problem has been studied like in Sect. \ref{sect_diff_mode}, as the limit of two distinct problems with values of $\lambda_{\rm pre}$
approaching 1 from below and above, respectively, being a periodic solution as the one in (\ref{var_cos_solSIX}) not admissible
when $\lambda_{\rm pre}=1$. The so-calculated buckling conditions for the two sets of electrostrictive materials are superposed in Fig. \ref{buckling}: part a) shows the dimensionless total stresses, while in part b) the dimensionless
electric displacement is pictured only for the electromechanical case.
Note that for  $k_1 h\ll 1$, for which the Eulerian instability theory is recovered, there is an outstanding
agreement between the buckling stresses for the cases of both mechanically  and electromechanically activated slab, while for higher $k_1 h$
appreciable differences arise.
Interestingly, the higher distance between the continuous curve and the scattered points of Fig. \ref{buckling}a pertains to the material with the higher degree of electrostriction:
this indicates that electrostriction strongly
influences the instability of the DE specimen, while a non-electrostrictive DE structure essentially buckles at a compressive stress very similar to that required in the purely mechanical case.


\section{Conclusions}

In soft dielectric elastomers, the electric permittivity may change considerably with the strain as a result of a strain-dependent
polarization response under an imposed electric field.
This phenomenon is called {\em electrostriction} and this paper addresses its modelling in the framework of
the general nonlinear theory of isotropic electroelasticity for both large and incremental deformations.
After having identified the relevant material parameters with experimental data,
in the second part of the article the general theory of bifurcation for electroelastic body proposed by Bertoldi and Gei (2011)
is applied to investigate mainly diffuse-mode bifurcations and band-localization instability, for which a detailed analysis is described
for the first time, showing that the theoretical condition for its existence
is only met when the dielectric solid displays an electrostrictive behaviour, being excluded otherwise.

Results show that electrostriction may activate the former modes at a threshold up to  30\% lower than that for an ideal dielectric (for which the
permittivity is constant), while we can argue that band-localization may trigger electric breakdown and failure of actual
prestretched/prestressed specimens. However, further theoretical and experimental
investigations are needed to clarify how localization may develop within a DE actuator under various electromechanical loading
conditions.

In the final part, a comparison between the buckling stresses of a mechanical compressed slab and the electrically activated
counterpart is performed, revealing that a high degree of electrostriction increases the critical stress and stiffens the layer
compared to the purely mechanical problem.

\vspace{10 mm}

{\bf Acknowledgements}. The financial supports of PRIN grant no. 2009XWLFKW, financed by Italian Ministry of Education, University and Research,
and of the COST Action MP1003 \lq European Scientific Network for Artificial Muscles', financed by EU, are gratefully acknowledged.

\begin{center}

{\bf Appendix A - Incremental constitutive moduli for the class of free energies represented by (\ref{iso_psi}).}\\

\end{center}

\paragraph{Total Lagrangian formulation.}
\beq
A^0_{MN}=\frac{1}{\epsilon_0 \bar \epsilon_r}\Big(\bar\gamma_0 \,\delta_{MN}+\bar\gamma_1\, C_{MN}+\bar\gamma_2\, C^2_{MN}\Big),
\eeq
\beq
\begin{split}
B^0_{iJM}&=\frac{1}{\epsilon_0 \bar \epsilon_r}\Big[\bar\gamma_1 (F_{iM}\, D^0_J +F_{iS}\, D^0_S\, \delta_{JM})+\\
&+\bar\gamma_2 (F_{iM}\, C_{JS}\,D^0_S+F_{iS}\, C_{JM}\,  D^0_S
+F_{iS}\, C_{SM}\, D^0_J+F_{iR}\, C_{RS}\, D^0_S\, \delta_{JM})\Big],
\end{split}
\eeq
\beqar
C^0_{iJkL}\!\! &= \!\!& \mu \Bigg[\bar\alpha_1 \delta_{ik} \delta_{JL} + 2 F_{iJ} F_{kL}
\left(\frac{\partial\bar\alpha_1}{\partial  I_1}- \bar\alpha_2\right) + \\
           & &   -2 \frac{\partial \bar\alpha_2}{\partial  I_2} (I_1 F_{iJ}-F_{iR} C_{RJ}) (I_1 F_{kL}-F_{kS} C_{SL})  + \nonumber   \\[2 mm]
           & & -  \bar\alpha_2 [ \delta_{ik}(I_1 \delta_{JL}-C_{JL})-F_{iL} F_{kJ}-B_{ik} \delta_{JL} ]       +  \nonumber \\[2 mm]
 & & +  2\frac{\partial\bar\alpha_1}{\partial  I_2} \left(2I_1 F_{iJ}F_{kL}-F_{iM}C_{MJ}F_{kL}-F_{iJ}F_{kM}C_{ML}\right)\Bigg] +\nonumber\\
           & & +\frac{1}{\epsilon_0 \bar \epsilon_r}\Big[\bar\gamma_1 \delta_{ik} D^0_J D^0_L+
           \bar\gamma_2 \left[ \delta_{ik} D^0_S (C_{JS}  D^0_L+C_{LS}  D^0_J) +\right.
           \nonumber   \\
&& +F_{iR} D^0_R (\delta_{JL} F_{kS} D^0_S+F_{kJ} D^0_L)
 \left. +F_{iL} F_{kS} D^0_S D^0_J+B_{ik} D^0_J D^0_L \right]\Big].\nonumber
\eeqar

\paragraph{Updated Lagrangian formulation.}
\beq
\lb{incr_moduli_updated_a}
A_{ab}=\frac{1}{\epsilon_0 \bar \epsilon_r} \Big(\bar{\gamma_0} B^{-1}_{ab}+\bar\gamma_1 \delta_{ab}+\bar\gamma_2 B_{ab}\Big),
\eeq
\beq
\lb{incr_moduli_updated_b}
B_{iqa}=\frac{1}{\epsilon_0 \bar \epsilon_r} \bigg[ \bar\gamma_1 (\delta_{ia} D_q +D_i \delta_{qa})+\bar \gamma_2 (\delta_{ia} B_{qs} D_s+ B_{qa}
D_i+ B_{ia} D_q+B_{is} D_s \delta_{qa})\bigg],
\eeq
%
\begin{eqnarray}
\lb{incr_moduli_updated_c}
C_{iqkp}\!\! &= \!\!&\mu \Bigg[ \bar \alpha_1 \delta_{ik} B_{pq} +2 B_{iq} B_{kp} \left( \derp{\bar \alpha_1}{I_1}-\bar \alpha_2 \right)+ \\
&~& -2 \derp{\bar \alpha_2}{I_2} \left( I_1 B_{iq}-B_{is}B_{sq} \right)\left( I_1 B_{kp}-B_{pt}B_{tk}\right)+\nonumber \\
&~& -\bar\alpha_2 (\delta_{ik} (I_1 B_{pq} -B_{pt}B_{tq})-B_{ip} B_{kq}-B_{ik} B_{pq} )+\nonumber \\
&~&2 \derp{\bar \alpha_1}{I_2}\left(2 I_1 B_{iq}B_{kp}-B_{is}B_{sq}B_{kp}- B_{iq}B_{ks} B_{sp}\right)\Bigg]  +  \nonumber \\
&~&  +\frac{1}{\epsilon_0 \bar \epsilon_r} \Bigg[ \bar \gamma_1 \delta_{ik} D_p D_q+
\bar \gamma_2 \bigg( \delta_{ik} D_s ( B_{qs} D_p+ B_{ps}D_q) + D_i (B_{pq} D_k+B_{qk} D_p) + \nonumber \\
&~& +D_q ( B_{pi}D_k+B_{ik} D_p) \bigg)\Bigg ].\nonumber
{}\end{eqnarray}



\end{document}